\documentclass[12pt]{article}
\pdfoutput=1

\usepackage[utf8]{inputenc}
\usepackage[left=2.55cm, right=2.55cm, top=2.55cm, bottom=2.55cm]{geometry}
\usepackage{amsmath,amssymb,amsbsy}
\usepackage{slashed}
\usepackage{xcolor}
\usepackage{graphicx}
\usepackage{url}
\usepackage{cancel}
\usepackage{cite}
\usepackage[colorlinks=true,allcolors=blue,pdfborder={0 0 0},linktocpage=false,pdfencoding=auto]{hyperref}
\usepackage{tabularx,booktabs}
\usepackage{multicol}
\usepackage{feynmp}
\usepackage{units}
\usepackage{xspace}
\usepackage[labelfont=bf]{caption}
\usepackage[section]{placeins}
\usepackage{subcaption}
\usepackage{soul} 
\usepackage{bbold}
\usepackage{parskip}
\usepackage{float}
\usepackage{multirow}

\definecolor{darkred}{rgb}{0.6,0,0}
\definecolor{darkpurple}{rgb}{0.5,0,0.5}

\newcommand{\code}[1]{\texttt{#1}}

\newcommand\snowmass{\begin{center}\rule[-0.2in]{\hsize}{0.01in}\\\rule{\hsize}{0.01in}\\
\vskip 0.1in Submitted to the  Proceedings of the US Community Study\\ 
on the Future of Particle Physics (Snowmass 2021)\\ 
\rule{\hsize}{0.01in}\\\rule[+0.2in]{\hsize}{0.01in} \end{center}}

\def\B0{B^{(0)}}
\def\A0{A_3^{(0)}}

\def\amu{$\Delta{a^{\rm FB}_\mu}$}

\begin{document}

\author{Amin Aboubrahim$^a$\footnote{\href{mailto:aabouibr@uni-muenster.de}{aabouibr@uni-muenster.de}}~, Michael Klasen$^a$\footnote{\href{mailto:michael.klasen@uni-muenster.de}{michael.klasen@uni-muenster.de}}~, Pran Nath$^b$\footnote{\href{mailto:p.nath@northeastern.edu}{p.nath@northeastern.edu}}~ and Raza M. Syed$^c$\footnote{\href{mailto:rsyed@aus.edu}{rsyed@aus.edu}} \\~\\
$^{a}$\textit{\normalsize Institut f\"ur Theoretische Physik, Westf\"alische Wilhelms-Universit\"at M\"unster,} \\
\textit{\normalsize Wilhelm-Klemm-Stra{\ss}e 9, 48149 M\"unster, Germany} \\
$^{b}$\textit{\normalsize Department of Physics, Northeastern University,
Boston, MA 02115-5000, USA} \\
$^{c}$\textit{\normalsize Department of Physics, American University of Sharjah, P.O. Box 26666, Sharjah, UAE} \\}

\title{\vspace{-3cm}\begin{flushright}
{\small MS-TP-21-26}
\end{flushright}
\vspace{0.5cm}
\Large{ \bf Future searches for SUSY at the LHC post  Fermilab $(g-2)_\mu$}
\vspace{0.5cm}}

\date{}
\maketitle

\vspace{-1cm}
\snowmass

\begin{abstract}
{
We assess the future directions for the search for supersymmetry at the Large Hadron Collider in view
of the new precision results on the muon anomaly by the Fermilab Collaboration. 
The existence of a deviation of size 4.1$\sigma$ from the Standard Model prediction points to light sleptons and light weakinos in the mass range of few hundred GeV while the observation of the 
Higgs boson mass at $\sim 125$ GeV points to squark masses lying in the few TeV range.
  Thus  a split sparticle spectrum is indicated. We discuss the possibility of such a split sparticle spectrum
  in the supergravity unified model and show that a splitting of the sfermion spectrum into light sleptons
  and heavy squarks naturally arises within radiative breaking of the electroweak symmetry driven by
  heavy gluinos ($\tilde g$SUGRA). We discuss the possible avenues for the discovery of supersymmetry 
  at the LHC within this framework under the further constraint of the recent muon anomaly result from the
  Fermilab  Collaboration. We show that the most likely candidates for early discovery of a sparticle 
  at the LHC are the chargino, the stau, the smuon and the selectron. We present a set of benchmarks 
  and   discuss future directions 
  for further work. Specifically, we point to
   the most promising 
  channels for SUSY discovery and estimate the integrated luminosity needed for the discovery of these benchmarks
  at the High Luminosity LHC and also at the High Energy LHC.
}
\end{abstract}

\numberwithin{equation}{section}

\newpage

{ \hypersetup{colorlinks=black,linktocpage=true} \tableofcontents }

\section{Introduction \label{1}}
 {
 
   The anomalous magnetic moment is one of the most precisely computed quantities in the Standard Model 
   and is predicted to have the value~\cite{Aoyama:2020ynm,Davier:2019can,Davier:2017zfy,Davier:2010nc,Crivellin:2020zul,Keshavarzi:2020bfy,Colangelo:2020lcg}   
 \begin{equation}
 a^{\rm SM}_\mu = 116 591 810 (43)  \times10^{-11}.
  \label{SM}
\end{equation}
The recent combined Fermilab~\cite{Abi:2021gix}  and Brookhaven~\cite{Bennett:2006fi,Tanabishi} measurement result gives 
 \begin{equation}
 a^{\rm exp}_\mu = 116 592 061 (41)  \times10^{-11}.
  \label{Brook}
\end{equation} 
The difference between the experimental and the Standard Model result is 
  \begin{equation}
\Delta a^{\rm FB}_\mu = a_{\mu}^{\rm exp} - a_{\mu}^{\rm SM} = 251 (59)  \times10^{-11}.
  \label{diff}
\end{equation} 
The above corresponds to a $4.1\sigma$ deviation from the Standard Model. This deviation strengthens
the previous Brookhaven observation of $3.7 \sigma$ (see, however,~\cite{Borsanyi:2020mff}).
The implications of these for supersymmetry in the framework of supergravity unified models
were analyzed in~\cite{Aboubrahim:2021rwz}. 
These works are motivated by the fact that it has been known since  the early eighties~\cite{Kosower:1983yw,Yuan:1984ww}  that a correction to the muon anomaly from supersymmetry can arise which is comparable to the
       electroweak correction in the Standard Model. Further, it is known that the muon anomaly is sensitive to
       soft parameters in supergravity models~\cite{Lopez:1993vi,Chattopadhyay:1995ae,Moroi:1995yh}.
        Thus the soft parameters directly affect the sparticle masses  that enter in the supersymmetric correction to the 
        muon anomaly, and specifically 
      when the
       smuon and muon-sneutrino as well as the chargino and the neutralino  are light with masses lying in the sub-TeV
       region, their loop corrections to the muon $g-2$ anomaly can be sizable. The above implies that within the framework of 
        high scale models, the $4.1\sigma$ deviation seen by Fermilab  can be explained provided we have
        light weakinos and light sleptons. On the other hand, the observation of the
         Higgs boson mass at $\sim 125$ GeV~\cite{Aad:2012tfa,Chatrchyan:2012ufa}         
          indicates
       that the size of weak scale supersymmetry lies in the several TeV region in order to generate a large
       enough loop correction to the Higgs boson mass to lift its tree value that lies below the $Z$-boson mass to the 
       experimentally observed value~\cite{Akula:2011aa,higgs7tev1}.       
       The supergravity based analysis discussed here proposes a natural
       solution to these apparently conflicting constraints on the model. Thus a possible solution to this
       disparity in the scale of the sfermion mass spectrum can come about if the radiative breaking of the
       electroweak symmetry in supergravity (SUGRA) grand unified models is driven by a large gluino mass lying in the
       several TeV region while
       the universal scalar mass is relatively small of order a few hundred GeV.      
    In this case squarks, which
     carry color, acquire large masses the size of several TeV in renormalization group (RG) running while the color neutral sleptons
      remain light down to the electroweak scale. 
    
    }
    
   In the numerical analysis we use an artificial neural network to search for regions of the parameter space consistent
           with the Fermilab result within the SUGRA parameter space and the search leads to the $\tilde g$SUGRA model
           proposed in earlier works~\cite{Akula:2013ioa,Aboubrahim:2019vjl,Aboubrahim:2020dqw}.          
        We will also discuss  a specific  $SO(10)$ 
           model, where a similar conclusion is reached~\cite{Aboubrahim:2021phn}.  
                    
\section{Fermilab $g-2$ constraint on SUGRA models}

 To discuss the Fermilab constraint on SUGRA models,  we utilize
 an artificial neural network to identify the region of the parameter space compatible with 
the muon anomaly result. Neural networks are efficient in analyses when one deals with a large
parameter space (see, e.g., Refs.~\cite{Hollingsworth:2021sii,Balazs:2021uhg}) and this is the 
case for models we investigate. 
Specifically we investigate 
the SUGRA model~\cite{sugrauni}
with non-universalities~\cite{Ellis:1985jn,nonuni2,Feldman:2009zc,Belyaev:2018vkl}  
 where the non-universalities are defined by the soft parameters consisting of the set $m_0, m_1,m_2,m_3, A_0, \tan\beta$ and sign$(\mu)$. Here 
$m_0$ defines the universal scalar mass for the sfermions and for the Higgs, $m_1, m_2, m_3$ are the
soft masses for the $U(1)$, $SU(2)$ and $SU(3)$ gauginos, $A_0$ is the universal scalar coupling all
taken at the grand unification scale. Further, $\tan\beta=\langle H_2\rangle/\langle H_1\rangle$ where $\langle H_2\rangle$ gives mass to 
the up quarks and $\langle H_1\rangle$ gives mass to the down quarks and the leptons.

 {As noted earlier there are apparent conflicting constraints that arise on the sparticle spectrum from the
  Higgs boson mass measurement at $\sim 125$ GeV and the muon $g-2$ anomaly measurement.
  The first one points to a large size of weak scale
  SUSY lying in the several TeV region while the muon anomaly result points to a relatively light scale
  for the sparticles in order to produce the desired correction to the anomaly in supersymmetry. Thus 
on the surface these two constraints appear to be at odds with each other. However, without any a priori assumption
 on the sizes of the soft parameters one can investigate regions of the parameter space which may lead to
 consistency using a neural network. An analysis within this framework shows that the preferred region within the 
 prescribed parameter space is one where the  
  universal scalar mass at the GUT scale is relatively small lying in the
    few hundred GeV region but the gluino mass is large and lying in the few TeV region.}
    
\begin{figure}[H]
  \begin{center}
    \includegraphics[scale=0.80]{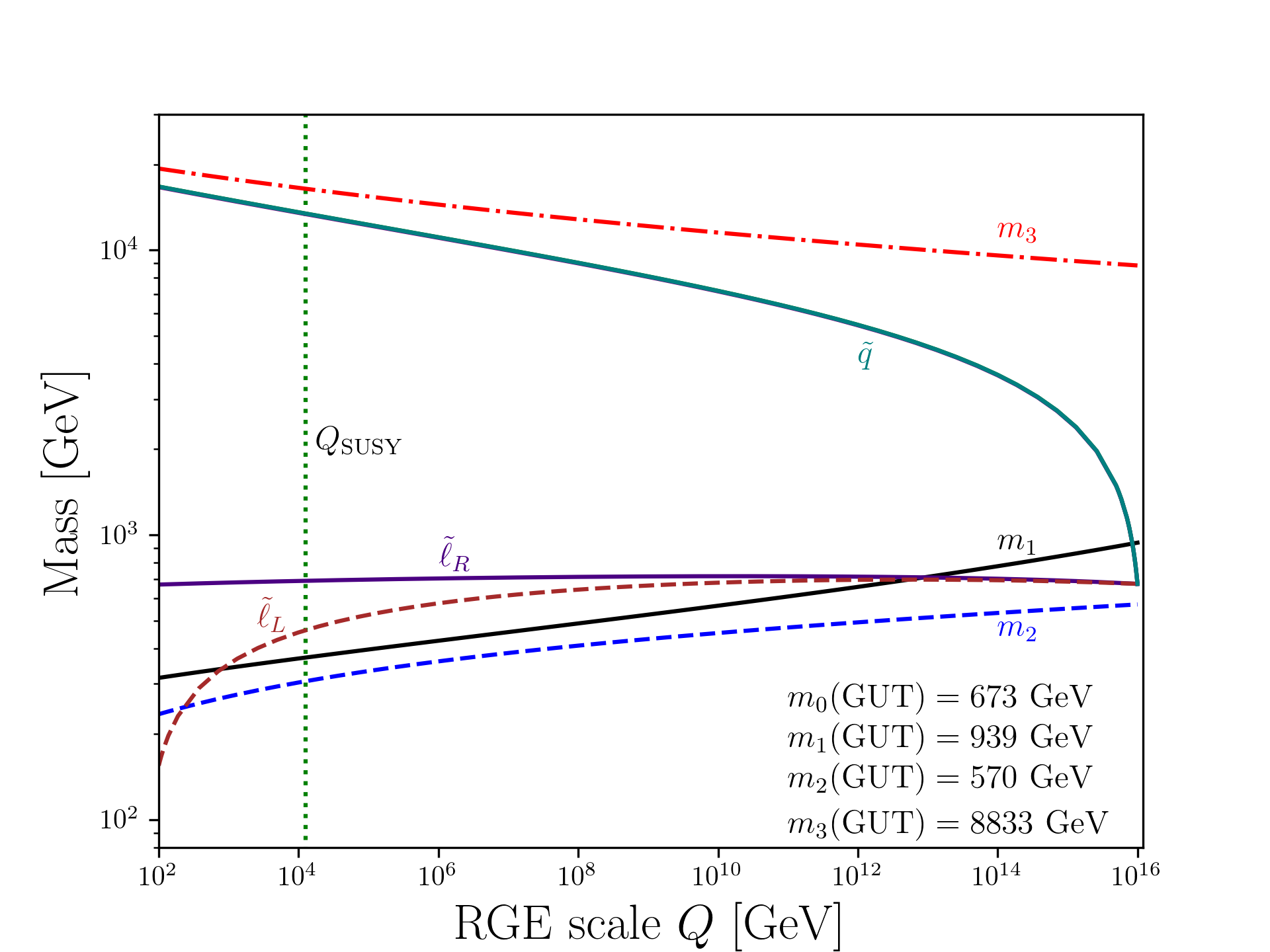}
    \caption{\label{splitting}
    Exhibition of a natural splitting between the slepton and squark masses in the gluino-driven radiative breaking
    of the electroweak symmetry. The running of the gaugino masses is also shown. The SUSY breaking scale, $Q_{\rm SUSY}$, is defined as the geometric average of the two stop masses. Based on the analysis of~\cite{Akula:2013ioa}.
   }
  \end{center}
\end{figure}

\begin{figure}[!htp]
\centering
\includegraphics[width=0.55\textwidth]{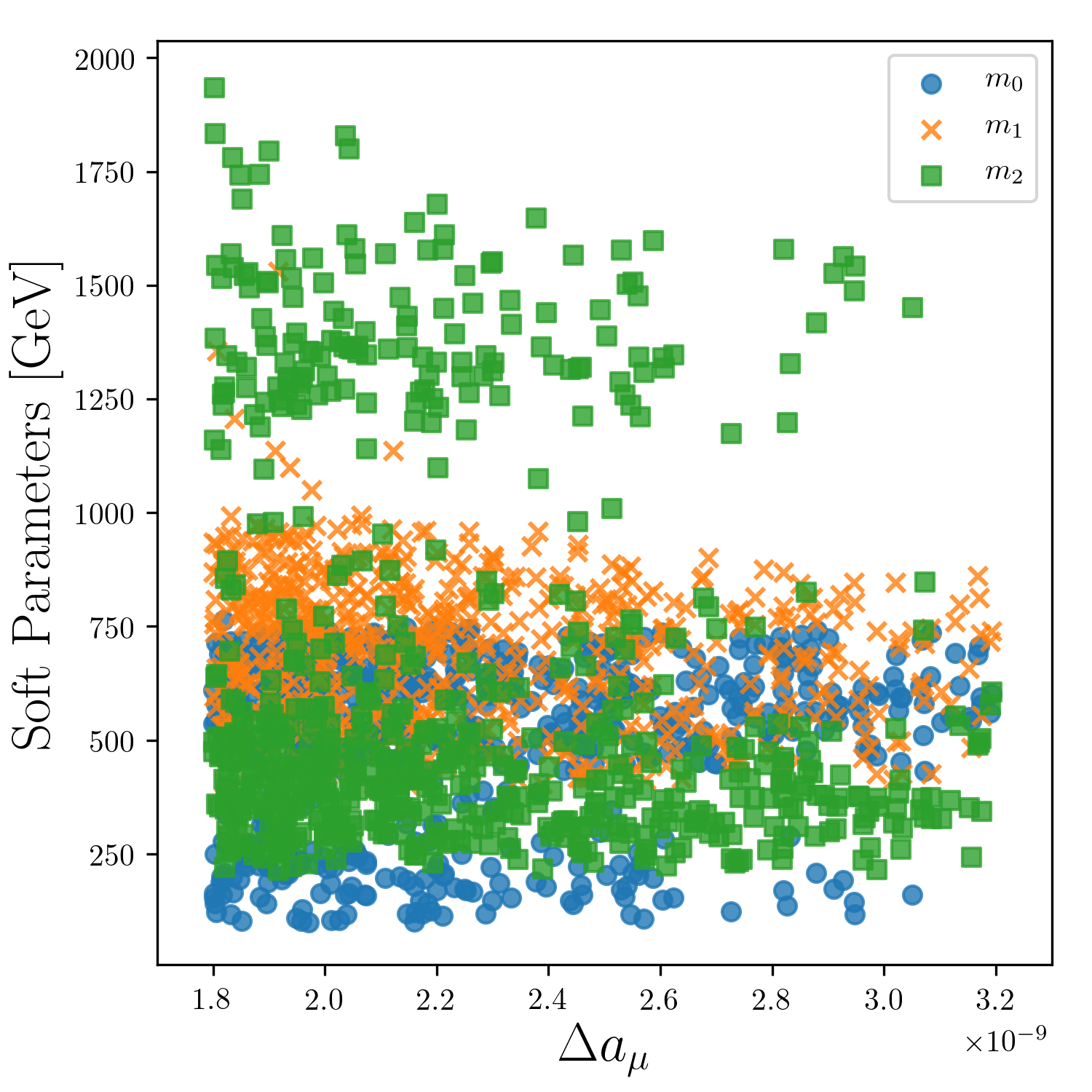}
\caption{A scatter plot of $m_0, m_1, m_2$ vs. $\Delta a_{\mu}$ using an artificial neural network analysis
consistent with the combined Fermilab-Brookhaven result within $1\sigma$.
The model points satisfy  the Higgs boson mass, the dark matter relic density and limits from dark matter direct detection experiments and the LHC
sparticle mass limits. }
\label{fig2}
\end{figure}

\begin{figure*}[t!]
  \begin{center}
    \includegraphics[scale=0.45]{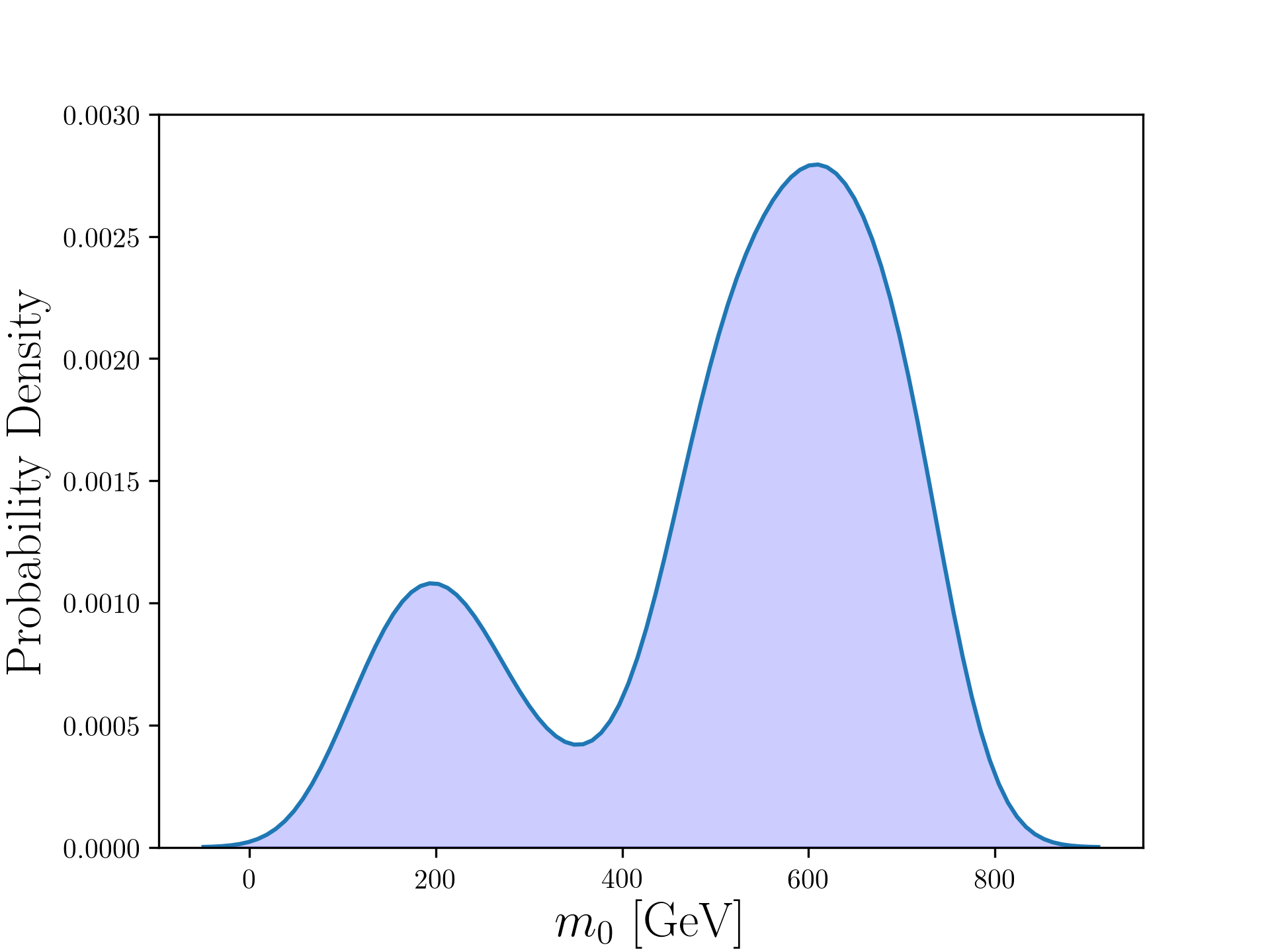}
    \includegraphics[scale=0.45]{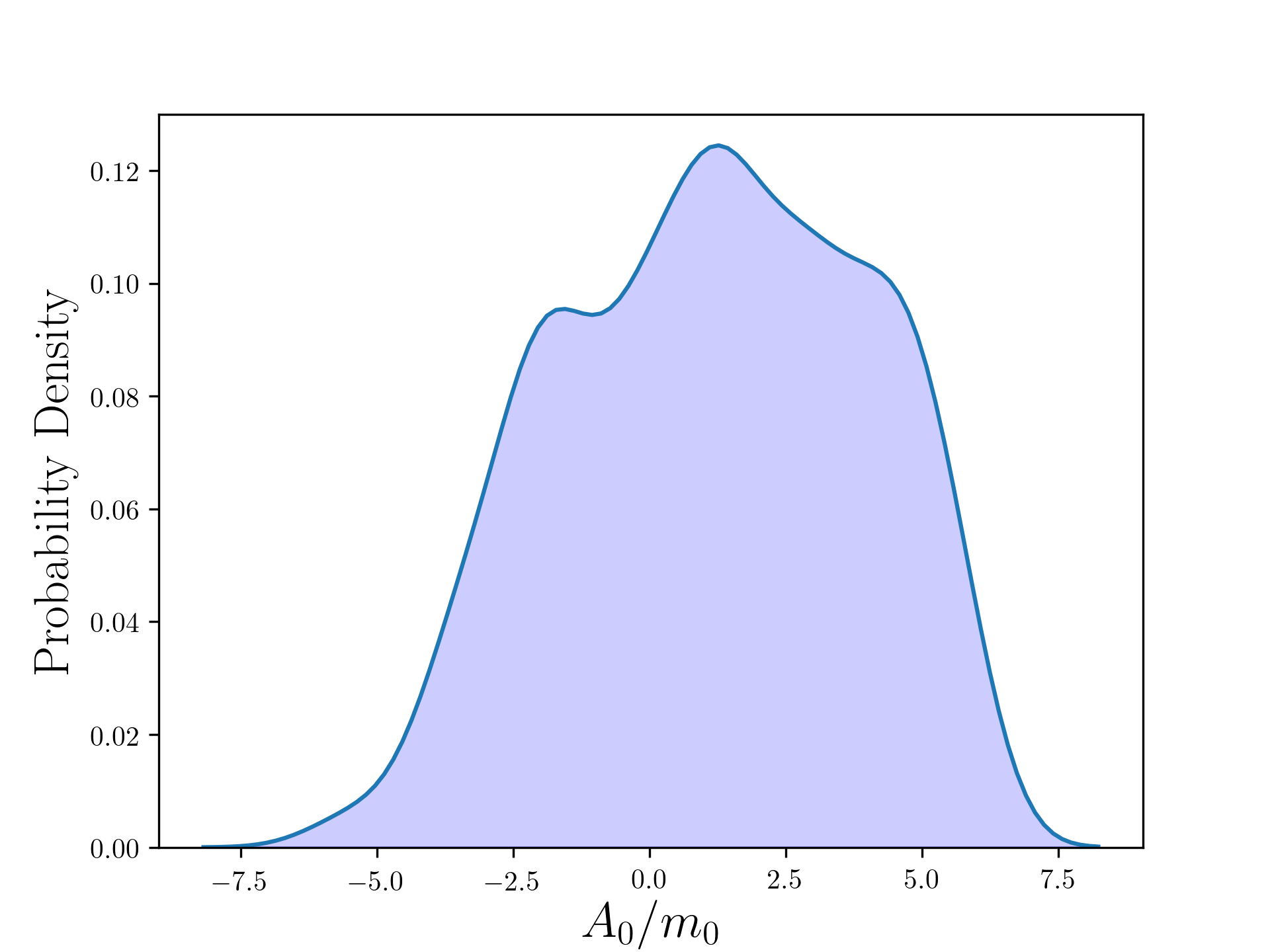} \\
     \includegraphics[scale=0.45]{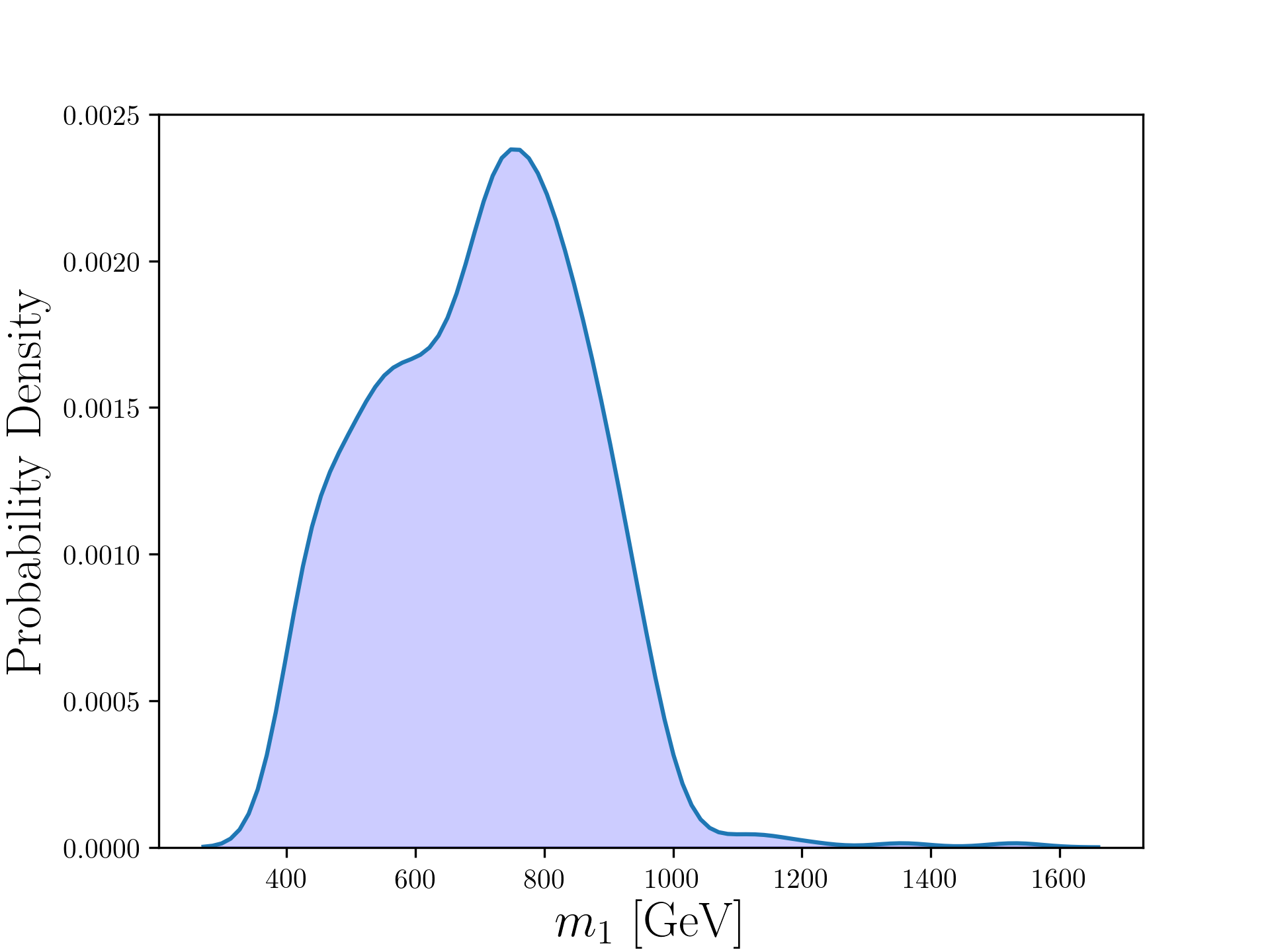}
    \includegraphics[scale=0.45]{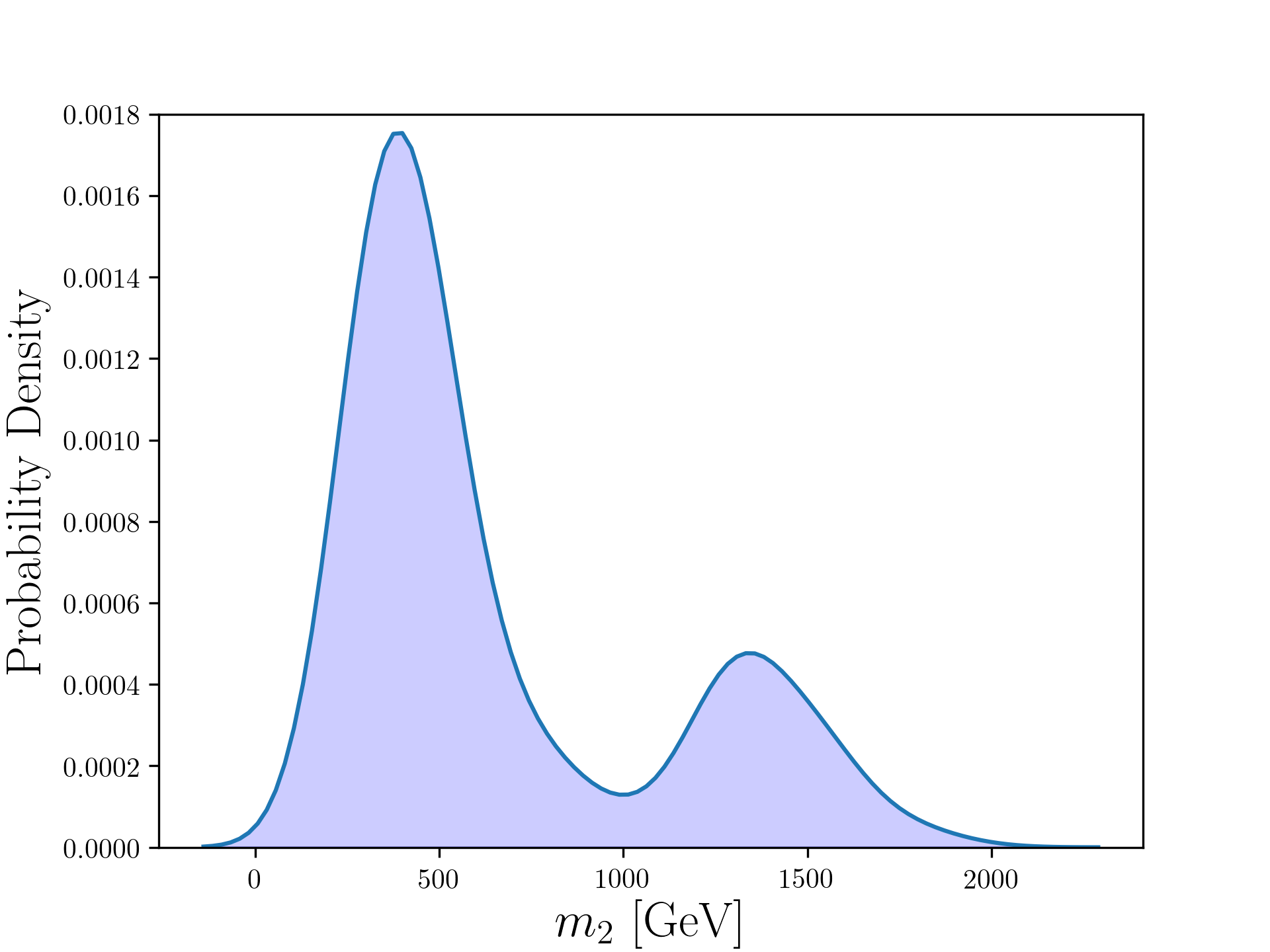}
    \caption{\label{a0tb}
    Probability density vs. the range of the soft parameters, $ m_0, A_0/m_0$ (top panels) and $m_1, m_2$ (bottom panels)     
     that emerge in the artificial neural network
     analysis of SUGRA parameter space which generates the desired correction to the muon anomaly indicated
     by the Fermilab experiment. }
  \end{center}
\end{figure*}

  As a consequence
     of the large gluino mass,  in the RG evolution the squark masses are driven to TeV scale due to their
    color interactions, while the color neutral  sleptons remain light. This mechanism then leads to
    a natural splitting between the slepton and squark masses as shown in Fig.~\ref{splitting} as noted in the
    early work of~\cite{Akula:2013ioa}.   
     The large mass of the squarks allows one
 to satisfy the Higgs boson mass constraint, while the relative smallness  of the slepton masses allows
 us to produce a significant correction to the muon $g-2$ anomaly indicated by the new measurements.
In Fig.~\ref{fig2} we give a  scatter plot of  SUGRA model points consistent with the Fermilab constraint
 with  plots of $m_0, m_1, m_2$ vs. $\Delta a_\mu$,  and in Fig.~\ref{a0tb}  we give the probability densities of $m_0$, $A_0/m_0$, $m_1$ and $m_2$  consistent with the $g-2$ constraint. 
  We note that in these plots   a significant population of models  exist within the $1\sigma$ range of
 $\Delta a^{\rm FB}_\mu$ given by Eq.~(\ref{diff}). Thus the satisfaction of the constraint occurs over
  a region of the SUGRA parameter space rather than just over few model points. Examining the distributions of Fig.~\ref{a0tb}, we notice an apparent anticorrelation between $m_0$ and $m_2$ and general broad distributions for $m_1$ and $A_0/m_0$. The reason behind this bimodal distribution in $m_0$ and $m_2$ is the LHC constraints. Limits on sparticle masses from LHC analyses disfavor small $m_0$ while a larger number of points at higher values of $m_0$ remain viable. The opposite is true for $m_2$ where larger $m_2$ values are disfavored. In particular, for $m_2>m_1$, the right handed slepton becomes lighter than the left handed one which becomes more constrained by LHC analyses.

\section{Sparticle mass hierarchy under the $g-2$ constraint}

\begin{figure*}[t]
\hspace{3cm} {Light Spectrum} \hspace{5.2cm} {Heavy Spectrum}
  \begin{center}
  \includegraphics[scale=0.64]{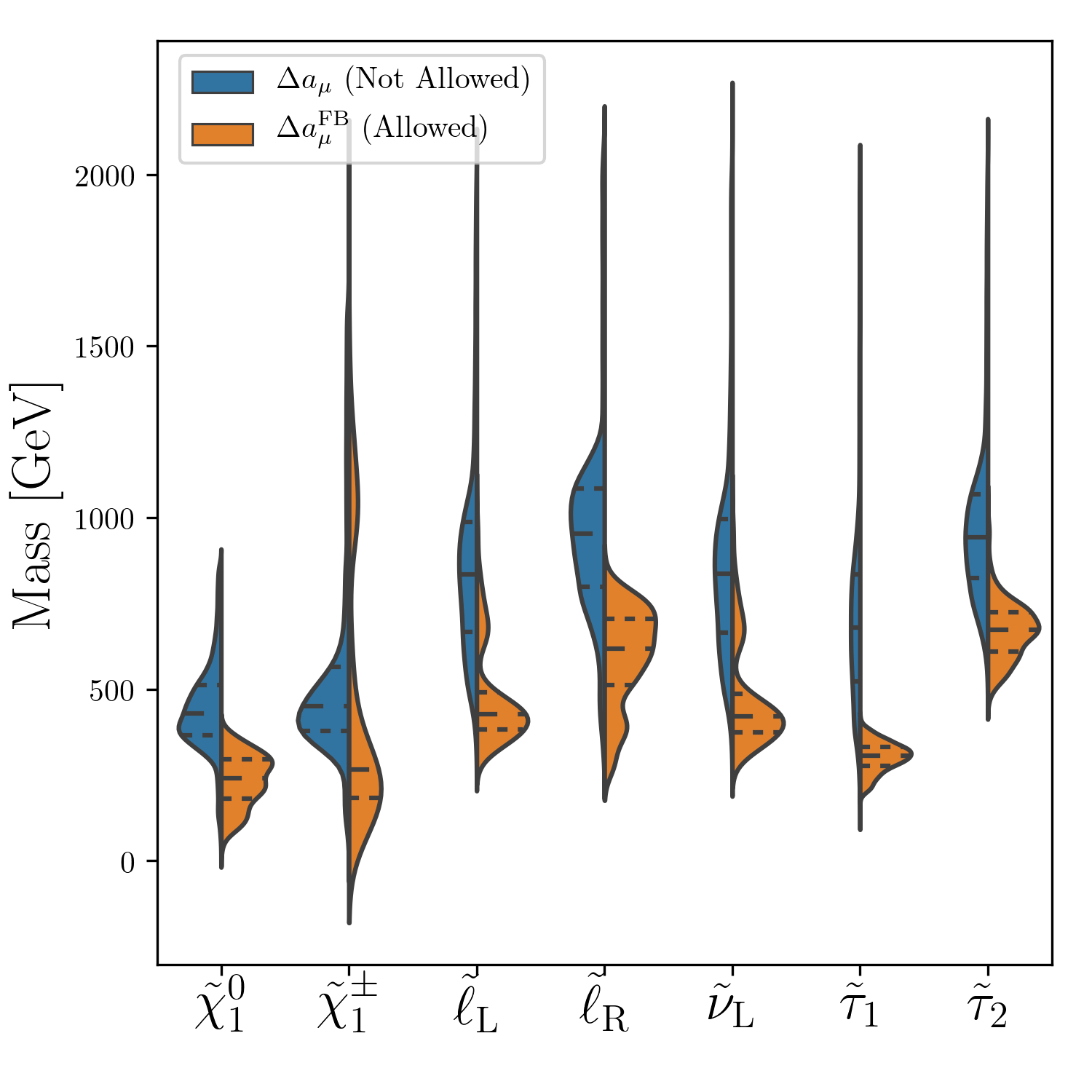}
  \includegraphics[scale=0.64]{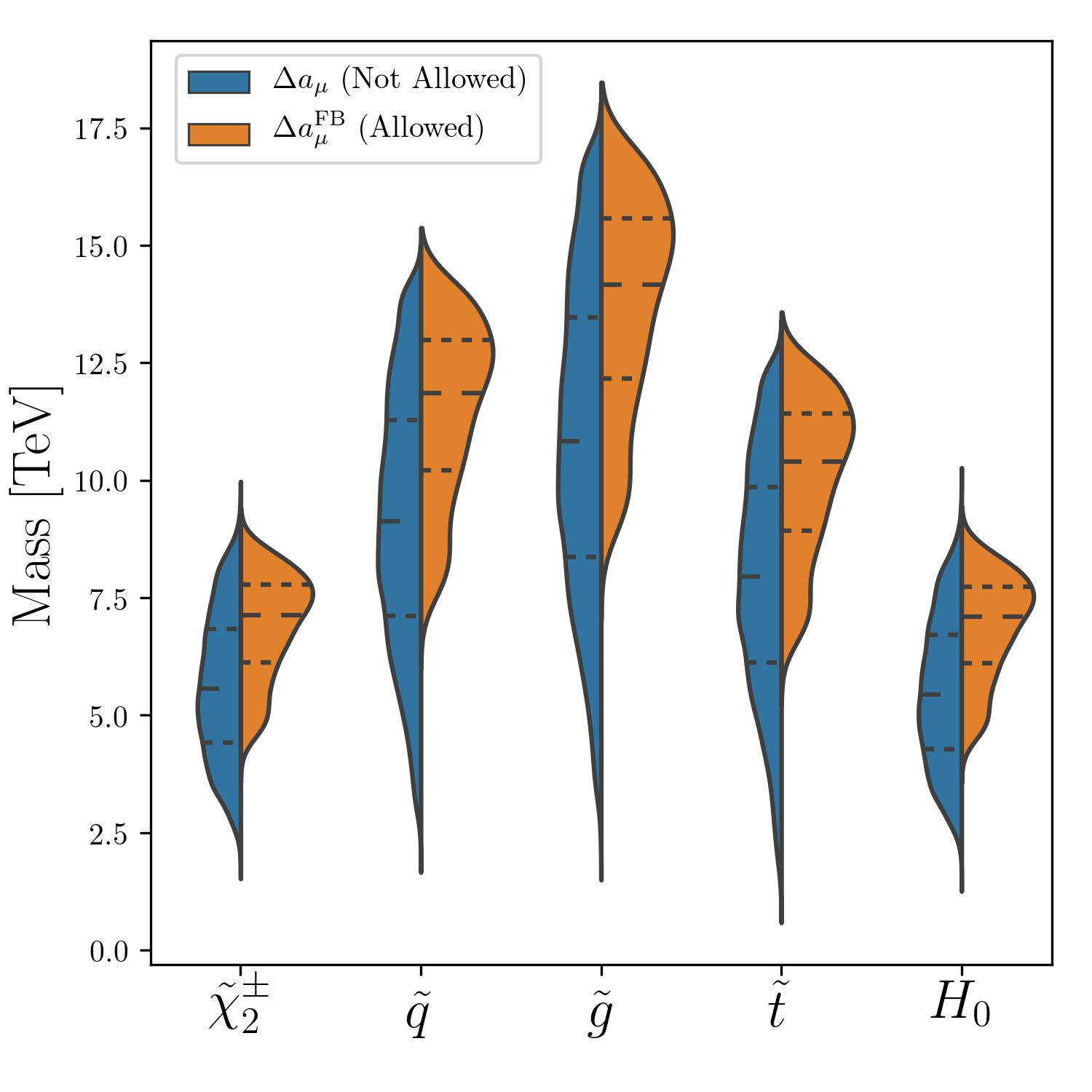}
    \caption{\label{splitmass}
   A display of the split sparticle spectrum consisting of light weakinos, sleptons, sneutrino and staus (left panel) and the heavy chargino, squarks, gluino, lightest stop and the CP even Higgs (right panel) that emerge in the
    gluino-driven radiative breaking of the electroweak symmetry in $\tilde g$SUGRA grand unified models. Each subplot shows a probability density distribution in the particles' masses for two cases: the region consistent with $\Delta a_{\mu}^{\rm FB}$ (orange) and  the region outside of the muon $g-2$ error bars (blue). } 
   \end{center}
\end{figure*}

As noted above  the satisfaction of the Higgs boson mass and 
the muon anomaly constraints leads to a mass hierarchy between the 
color neutral  sfermions and the sfermions with color as seen from Fig.~\ref{splitting}. A priori, a large mass gap between the sleptons and the squarks appears puzzling. This is so since 
sleptons and squarks belong to common multiplets in grand unification, e.g., a single generation
of leptons and quarks and their superpartners lie in one 16-plet representation of $SO(10)$, and
soft breaking does not distinguish between colors. Thus the remarkable aspect of $\tilde g$SUGRA
is that renormalization group naturally leads to a split spectrum between sleptons and squarks,  
  because the squarks
  couple to the gluino at the loop level and the large mass of the gluino thus drives their masses to large
  values while sleptons are largely unaffected.  A display of the sparticle mass hierarchy 
  generated by the RG running is given in Fig.~\ref{splitmass}. Here the left panel shows the
  light sparticle spectrum while the right panel shows the heavy sparticle masses. Corrections to the muon
  $g-2$ are governed by the light spectrum of the left panel and this mass spectrum lies in the mass
  range of few hundred GeV while the heavy mass spectrum lies in the mass range of few TeV.
  Because of this split mass spectrum, the sparticles most accessible for SUSY searches at the LHC
  are the ones in the left panel.  For the low-lying spectrum consistent with $\Delta a_{\mu}^{\rm FB}$, the mean value (shown by the dashed line) of the chargino mass is $\sim 445$ GeV,  the lightest neutralino mass $\sim 235$ GeV, the left handed slepton mass $\sim 480$ GeV, the right handed slepton mass $\sim 590$ GeV, the sneutrino mass $\sim 470$ GeV and two stau mass eigenstates, $\tilde\tau_1$ and $\tilde\tau_2$, are $\sim 300$ GeV and $\sim 670$ GeV, respectively. The heavy spectrum includes the second chargino, the squarks, the gluino, the lightest stop and the CP even Higgs $H_0$. Note that the charged and CP odd Higgses have masses comparable to $H_0$. The dashed lines on either side of the mean correspond to the quartiles of this distribution.

It is also of interest to investigate the effect of the $g-2$ constraint on specific grand unified models.
As a model example, we consider the case of an $SO(10)$ model where the doublet-triplet 
 mass splitting is obtained in a natural fashion via the missing partner mechanism. For $SU(5)$ 
  this is accomplished in~\cite{Masiero:1982fe,Grinstein:1982um} while for $SO(10)$ in~\cite{Babu:2006nf,Babu:2011tw,Aboubrahim:2020dqw,Aboubrahim:2021phn} (for a review of 
  other grand unified models see~\cite{Nath:2006ut}).
The $SO(10)$ model also allows for a $b-t-\tau$ unification~\cite{Ananthanarayan:1992cd,Chattopadhyay:2001va} along with other desirable properties.
  A neural network analysis again leads us to conclusions similar to the above, i.e., that the radiative electroweak symmetry breaking is gluino driven. Further, we note that in the analyses of  the cases discussed above, the light sparticle spectrum falls in three
  classes of mass hierarchies which we label as (A), (B) and (C)  
  as  given below.

\paragraph{(A):}  Here
$\tilde\chi^0_2,\tilde\chi^{\pm}_1$ are essentially degenerate, and $\tilde \tau_1$ is the next lightest 
  supersymmetric particle (NLSP). This leads to the mass hierarchy 
$$m_{\tilde\tau_1}<m_{\tilde\chi^0_2}, 
m_{\tilde\chi^{\pm}_1}<m_{\tilde\ell},$$
where $\tilde\ell$ stands for selectron or smuon.
\paragraph{(B):} Here there is a reversal in the hierarchy for the first two inequalities, i.e., between $\tilde \tau_1$
and  $\tilde\chi^0_2$ or $\tilde\chi^{\pm}_1$ which leads to the following possibilities 
\begin{align*}
&m_{\tilde\chi^0_2},  m_{\tilde\chi^{\pm}_1}<m_{\tilde\tau_1}<m_{\tilde\ell},\\
&m_{\tilde\chi^{\pm}_1}<m_{\tilde\chi^0_2}<m_{\tilde\tau_1}<m_{\tilde\ell},\\
&m_{\tilde\chi^{\pm}_1}<m_{\tilde\tau_1}<m_{\tilde\chi^0_2}<m_{\tilde\ell}.
\end{align*}
\paragraph{(C):} In this case the selectron and the smuon are lighter than the chargino and the second neutralino 
while the stau is the NLSP. Thus here we have
$$m_{\tilde\tau_1}<m_{\tilde\ell}< m_{\tilde\chi^0_2},  m_{\tilde\chi^{\pm}_1}.$$

\begin{figure}[H]
\centering
\includegraphics[width=0.75\textwidth]{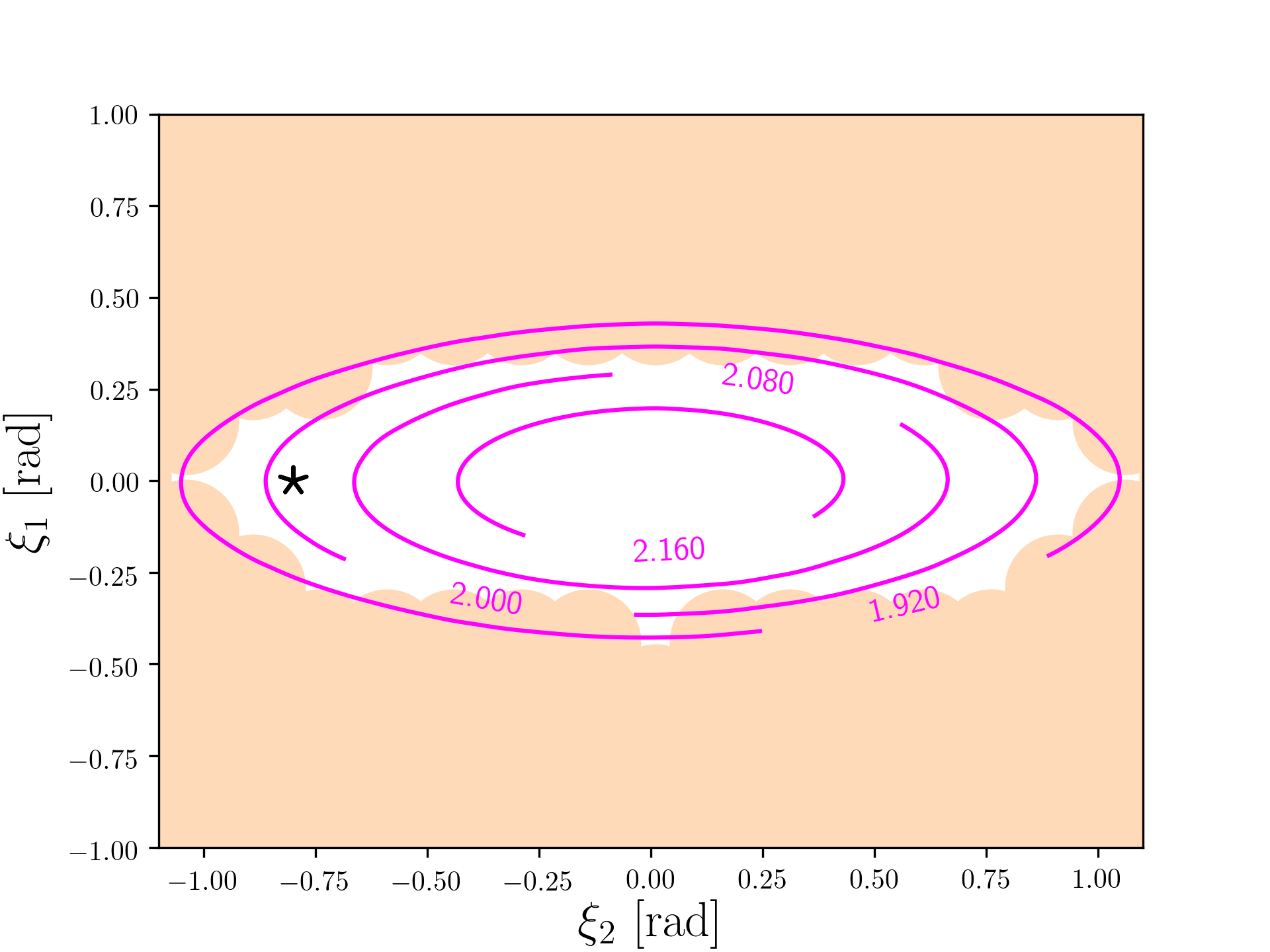}
\caption{
A display of the region that
is excluded (shaded) and the region that is allowed (white) by the $g_\mu-2$ constraint.
 The allowed region is further constrained
by the limits on the EDM of the electron and the EDM of the neutron. The star in the white region represents the small region of the parameter space where both the electron EDM and the muon $g-2$ are satisfied. Based on analysis of~\cite{Aboubrahim:2021rwz}.}
\label{figcp}
\end{figure}

The Fermilab result also puts constraints on the  allowed region of CP phases arising from the
soft parameters specifically the gaugino masses. Thus the gaugino masses $m_1$ and $m_2$ 
can be complex and one may write $m_1=|m_1| e^{i\xi_1}$ and $m_2= |m_2|e^{i\xi_2}$.
It turns out that the electroweak corrections to the muon $g-2$ are very sensitive to the CP phases~\cite{Ibrahim:1999aj,Ibrahim:2001ym,Aboubrahim:2016xuz}.
As a consequence the measured \amu~puts rather stringent constraints on the allowed range of 
 the CP phases. This is shown in  Fig.~\ref{figcp} which exhibits the disallowed region (shaded) and
 the allowed region (white) in the $\xi_1-\xi_2$ plane. It is to be noted that the allowed region is further
 constrained by the experimental limits on the EDM of the electron and of the neutron. Thus the allowed region resulting from imposing both constraints becomes very narrow. We label this region by a star in Fig.~\ref{figcp} which can be obtained via the cancellation mechanism~\cite{Ibrahim:1998je}. The CP phases
 can also affect dark matter analyses (see, e.g.,~\cite{Chattopadhyay:1998wb}) and proton stability~\cite{Ibrahim:2000tx} as well as the production cross sections and decays of the sparticles.
 These effects are also worth investigation in the future in view of the current experimental result on
 the possible deviation \amu~from the Standard Model prediction.
 
\section{Benchmarks for future SUSY searches at the LHC consistent with $g-2$}

As noted earlier, the Fermilab $g-2$ measurement has important implications for the discovery of supersymmetry and future searches at colliders and elsewhere. Here we present 
 a set of SUGRA benchmarks consistent with the recent muon $g-2$ anomaly result to test for a potential SUSY discovery at HL-LHC and HE-LHC. The benchmarks are given in Table~\ref{tab1} and arranged according to the categories (A), (B) and (C) described earlier. Thus benchmark (a) belongs to case (A), benchmarks (b), (c), (d) belong to case (B) and benchmarks (e), (f), (g) belong to case (C). The benchmarks have an $m_0$ of $\mathcal{O}(100)$ GeV while  $m_3$ lies in 
   the several TeV range consistent with a gluino-driven radiative electroweak symmetry breaking. It is noted that benchmark (b) is taken from Ref.~\cite{Aboubrahim:2021rwz} (labeled there as (d)) while (a), (c), (d), (e), (f), (g) are from Ref.~\cite{Aboubrahim:2021phn} (labeled there as (f), (i), (b), (h), (g), (d), respectively). 

\begin{table}[H]
\caption{\label{tab1}
SUGRA benchmarks for future SUSY searches. All masses are in GeV.}
\centering
\begin{tabular}{|ccccccc|}
\hline\hline
Model & $m_0$ & $A_0$ & $m_1$ & $m_2$ & $m_3$ & $\tan\beta$ \\
   \hline\rule{0pt}{3ex}
\!\!(a) & 688 & 1450 & 852 & 634 & 8438 & 16.8 \\
(b) & 389 & 122 & 649 & 377 & 4553 & 8.2 \\ 
(c) & 452 & 648 & 624 & 346 & 4843 & 13.1\\
(d) & 673 & 1127 & 939 & 570 & 8833 & 8.2  \\
(e) & 206 & 603 & 842 & 1298 & 7510 & 8.0 \\
(f) & 106 & 22.6 & 523 & 1309 & 5240 & 9.3 \\
(g) & 164 & 197 & 632 & 1539 & 6171 & 12.2  \\
\hline\hline 
\end{tabular}
\end{table}
{
Along with the satisfaction of the recent muon $g-2$ result, the benchmarks satisfy the Higgs boson mass, the dark matter relic density as well as the LHC mass limits constraints. 
We present in Table~\ref{tab2} the low-lying sparticle spectrum along with the SM-like Higgs boson mass, the relic density and the muon $g-2$ anomaly calculated at the two-loop level. Here the left handed slepton is lighter than the right handed one except in benchmarks (e), (f) and (g) due to $m_2$ being significantly greater that $m_1$ as seen in Table~\ref{tab1}. Furthermore, the benchmarks respect current constraints on the proton-neutralino spin-independent cross section and all allow for a multicomponent dark matter scenario.  
}

\begin{table}[H]
\centering
\caption{\label{tab2}
The SM-like Higgs mass, the light sparticle spectrum and the dark matter relic density $\Omega h^2$ for the benchmarks of Table \ref{tab1}. Also shown is the muon $g-2$. }
\begin{tabular}{|cccccccccc|}
\hline\hline
Model & $h^0$  & $\tilde\ell_{\rm L}$ & $\tilde\ell_{\rm R}$ & $\tilde\nu_{\rm L}$ & $\tilde\tau_1$ & $\tilde\chi^0_1$ & $\tilde\chi^{\pm}_1$ & $\Omega h^2$ & $\Delta a_{\mu}(\times 10^{-9})$ \\
\hline\rule{0pt}{3ex}
\!\!(a) & 123.0 & 508.1 & 762.0 & 502.3 & 331.9 & 324.2 & 404.3 & 0.004 & 2.11 \\
(b) & 123.4 & 305.0 & 463.0 & 295 & 251.7 & 237.4 & 237.6 & 0.002 & 2.33 \\
(c) & 123.7 & 346.8 & 511.9 & 338.0 & 240.3 & 205.6 & 205.8 & 0.001 & 2.67 \\
(d) & 125.3 & 422.8 & 763.8 & 415.7  & 370.4  & 337.3 &  337.6 & 0.003 & 2.14 \\
(e) & 124.5 & 628.7 & 402.2 &  623.6  & 338.3 & 326.8 & 998.4  & 0.082 & 1.94 \\
(f) & 123.4 & 722.8 &  262.9 & 718.2  & 206.5 & 195.5 & 1038.4 &  0.103 & 2.57 \\
(g) & 123.9 & 856.4 & 327.4 & 852.4 & 243.5 & 240.1 & 1227  & 0.016 & 1.94\\
\hline\hline
\end{tabular}
\end{table}

We investigate three main SUSY production channels at the LHC: slepton and sneutrino pair production and slepton associated production with a sneutrino. The latter has a significant cross section since it proceeds via the charged current. We calculate the production cross sections at 14 TeV and 27 TeV at the aNNLO+NNLL accuracy using \code{Resummino-3.0}~\cite{Debove:2011xj,Fuks:2013vua}. The results are displayed in Table~\ref{tab3}. Notice that the contribution from the right handed sleptons becomes important in benchmarks (e), (f) and (g) where $m_{\tilde\ell_{\rm R}}<m_{\tilde\ell_{\rm L}}$.

\begin{table}[H]
\centering
\caption{\label{tab3}
The aNNLO+NNLL production cross-sections of slepton pair $(\tilde \ell= \tilde e, \tilde \mu)$ and sneutrino pair as well as slepton associated production with a sneutrino  at $\sqrt{s}=14$ TeV and $\sqrt{s}=27$ TeV for the benchmarks of 
Table \ref{tab1}. The cross section is in fb.} 
\begin{tabular}{|ccc|cc|cc|cc|}
\hline\hline
Model & \multicolumn{2}{c|}{$\sigma(pp\rightarrow \tilde\ell_{\rm L}\,\tilde\ell_{\rm L})$} & \multicolumn{2}{c|}{$\sigma(pp\rightarrow \tilde\ell_{\rm R}\,\tilde\ell_{\rm R})$} & \multicolumn{2}{c|}{$\sigma(pp\rightarrow \tilde\nu_{\rm L}\,\tilde\ell_{\rm L})$}& \multicolumn{2}{c|}{$\sigma(pp\rightarrow \tilde\nu_{\rm L}\,\tilde\nu_{\rm L})$} \\
\hline
&14 TeV & 27 TeV & 14 TeV & 27 TeV & 14 TeV & 27 TeV & 14 TeV & 27 TeV \\
\hline \rule{0pt}{3ex}
\!\!(a) & 1.084 & 4.515 & 0.057 & 0.335 &   &   &  &    \\
(b) & 9.810 & 30.80 & 0.656 & 2.54 & 41.78 & 127.80 & 10.52 & 33.10 \\
(c) & 5.805 & 19.31 & 0.417 & 1.72 & 24.40 & 78.97 & 4.94 & 15.10 \\
(d) & 2.486 & 9.188 & 0.056 & 0.332 & 10.44 & 37.47 & 2.10 & 7.21 \\
(e) & 0.388 & 1.915 & 1.22 & 4.32  &   &  &  &  \\
(f) & 0.188 & 1.065 & 6.91 & 20.15 &  &  &  &  \\
(g) & 0.074 & 0.506 & 2.87 & 9.22 &  &  &  &   \\
  \hline
\end{tabular}
\end{table}

In LHC analyses with simplified models, unit branching ratios for specific production channels are often used.
However, this simplified assumption does not hold in high scale models where the branching ratios exhibit a rich diversity of final states.
 Thus the  exclusion limits on sparticle masses do not apply directly to high scale models, but rather limits on $\sigma\times\text{BR}$ become the relevant ones. To illustrate this we display the different decay channels of the sleptons and sneutrinos 
 in Table~\ref{tab4}.

\begin{table}[H]
\centering
\caption{\label{tab4}
The branching ratios of relevant decay channels of the left and the right handed slepton and the sneutrino for the benchmarks of Table \ref{tab1}. } 
\begin{tabular}{|ccccccc|}
\hline\hline\rule{0pt}{3ex}
Model & $\tilde\ell_{\rm L}\to \ell\tilde\chi^0_1$ & $\tilde\ell_{\rm L}\to \ell\tilde\chi^0_2$ & $\tilde\ell_{\rm L}\to \nu_{\ell}\tilde\chi^{\pm}_1$ & $\tilde\ell_{\rm R}\to \ell\tilde\chi^0_1 [\tilde\chi^0_2]$ & $\tilde\nu_{\rm L}\to \tilde\chi^+_1\ell^-$ & $\tilde\nu_{\rm L}\to \tilde\chi^0_1\nu_\ell$ \\
\hline \rule{0pt}{3ex}
\!\!(a) & 22\% & 26\% & 52\% & 100\% [-] & 51\% & 24\%  \\
(b) & 31\% & 7\% & 62\% & - [100\%] & 62\% & 30\%  \\
(c) & 31\% & 6\% & 63\% & - [100\%] & 62\% & 31\% \\
(d) & 31\% & 6\% & 63\% & - [100\%] & 63\% & 32\%  \\
(e) & 100\% & - & - & 100\% [-] & - & 100\% \\
(f) & 100\% & - & - & 100\% [-] & - & 100\% \\
(g) & 100\% & - & - & 100\% [-] & - & 100\% \\
  \hline
\end{tabular}
\end{table}

\section{A deep neural network analysis for SUSY discovery}

{As can be seen from Table~\ref{tab4}, there are numerous decay channels for the sleptons and sneutrinos which result in rich final states. Leptonic channels are the most prevalent and for these we consider final states consisting of two leptons of the same flavour and opposite sign (SFOS) and missing transverse energy (MET) due to the neutralinos (and neutrinos). We design two signal regions, where in one we require exactly one non-b-tagged jet and in the other we require at least two non-b-tagged jets. One can easily see how slepton pair production can give us exactly such final states but it may be less trivial for sneutrino pair production and for slepton-sneutrino associated production. In Table~\ref{tab3} the cross section for sneutrino pair production and slepton associated production for benchmarks (b), (c) and (d) are shown. 
As mentioned earlier, those benchmarks belong to case (B) where the chargino is nearly mass degenerate with the LSP. 
Thus it is not easy to distinguish between these particles at the LHC and so we assume that the chargino escapes undetected and adds to the MET in an event. With a 62\% branching ratio for $\tilde\nu_{\rm L}\to\tilde\chi^+_1\ell^-$ (see Table~\ref{tab4}), sneutrino pair production will result in the same final states with a significant $\sigma\times\text{BR}$. The same applies for slepton associated production with a sneutrino which has the largest cross section among the different production channels considered. The rest of the benchmarks do not have a degenerate chargino-LSP and so the decay of the chargino in those benchmarks will result in very different final states.  }

{The signal, which  involves the different production channels presented in Table~\ref{tab3}, and the Standard Model background events are simulated at LO with \code{MadGraph5}~\cite{Alwall:2014hca} and showered with \code{PYTHIA8}~\cite{Sjostrand:2014zea} with the addition of ISR and FSR jets. Cross sections are scaled to their NLO values for the background and to aNNLO+NNLL for the signal. Detector effects are added by the help of \code{DELPHES-3.4.2}~\cite{deFavereau:2013fsa}.  
  }
  
{  
The analysis presented here shows that a discrimination between the signal and background events can be 
done using deep neural network (DNN) as part of the `Toolkit for Multivariate Analysis' (TMVA)~\cite{Speckmayer:2010zz} framework within \code{ROOT6}~\cite{Antcheva:2011zz}. Thus the possibility of detection of  light sparticles appears strong at HL-LHC and with much greater certainty at HE-LHC.

We use a set of kinematic variables to train the DNN on the signal and background events. The list of variables include
\begin{align*}
&E^{\rm miss}_{\rm T},~~~p_{\rm T}(j_1),~~~p_{\rm T}(\ell_1),~~~p_{\rm T}(\ell_2),~~~p_{\rm T}^{\rm ISR},~~~M_{\rm T2},~~~M^{\rm min}_{\rm T},~~~m_{\ell\ell}, \\
&\hspace{2.5cm}~~~\Delta\phi(\textbf{p}_{\rm T}^{\ell},\textbf{p}^{\rm miss}_{\rm T}),~~~\Delta\phi_{\rm min}(\textbf{p}_{\rm T}(j_i),\textbf{p}^{\rm miss}_{\rm T}), 
\end{align*}
where $M^{\rm min}_{\rm T}=\text{min}[m_{\rm T}(\textbf{p}_{\rm T}^{\ell_1},\textbf{p}^{\rm miss}_{\rm T}),m_{\rm T}(\textbf{p}_{\rm T}^{\ell_2},\textbf{p}^{\rm miss}_{\rm T})]$, $\Delta\phi(\textbf{p}_{\rm T}^{\ell},\textbf{p}^{\rm miss}_{\rm T})$ is the opening angle between the MET system and the dilepton system and $\Delta\phi_{\rm min}(\textbf{p}_{\rm T}(j_i),\textbf{p}^{\rm miss}_{\rm T})_{i=1,2,3}$ is the smallest opening angle between the first three leading jets in an event and the MET system. Note that $\textbf{p}_{\rm T}^{\ell}=\textbf{p}_{\rm T}^{\ell_1}+\textbf{p}_{\rm T}^{\ell_2}$. After reconstructing the momentum of the dilepton system, we determine the angle between the dilepton system and each non-b-tagged jet in the event. We select up to two jets that are closest to the dilepton system and tag them as possible jets arising from the decay of the SUSY system while the rest are classified as ISR jets. Here $p_{\rm T}^{\rm ISR}$ denotes the total transverse momentum of all the ISR jets in an event.  
}

\begin{table}[H]
\caption{The preselection criteria and the analysis cuts on a set of kinematic variables at 14 TeV (27 TeV) grouped by the benchmarks of Table~\ref{tab1} in two signal regions SR-$2\ell1$j and SR-$2\ell2$j. Entries with a dash (-) imply  that no requirement on the variable is considered.  Cuts are optimized for each center-of-mass energy. }
\resizebox{\textwidth}{!}{\begin{tabular}{|c|ccc|ccc|}
\hline\hline
\multirow{2}{*}{Observable}  & (b), (c), (d) & (a) & (e), (f), (g)  & (b), (c), (d) & (a) & (e), (f), (g)  \\
\cline{2-7}\rule{0pt}{3ex}
 & \multicolumn{3}{c|}{Preselection criteria (SR-$2\ell2$j)} & \multicolumn{3}{c|}{Preselection criteria (SR-$2\ell1$j)} \\
 \hline \rule{0pt}{3ex}
 $N_{\ell}$ (SFOS) & \multicolumn{3}{c|}{$2$} & \multicolumn{3}{c|}{$2$} \\
$N_{\rm jets}^{\rm non-b-tagged}$ & \multicolumn{3}{c|}{$\geq 2$} & \multicolumn{3}{c|}{$1$}  \\
$p_T(j_1)$ [GeV] & \multicolumn{3}{c|}{$>20$} & \multicolumn{3}{c|}{$>20$} \\
$p_T(\ell_1)$ (electron, muon) [GeV] & \multicolumn{3}{c|}{$>15$, $>10$} & \multicolumn{3}{c|}{$>15$, $>10$} \\
$E^{\rm miss}_T$ [GeV] & \multicolumn{3}{c|}{$>100$} & \multicolumn{3}{c|}{$>100$}\\
\cline{2-7}\rule{0pt}{3ex}
 & \multicolumn{3}{c|}{Analysis cuts} & \multicolumn{3}{c|}{Analysis cuts} \\
\cline{2-7}\rule{0pt}{3ex}
\!\! $m_{\ell\ell}~\text{[GeV]} >$ & 136 (110) & 150  & 150 (110) & 110 & 120  & 150   \\
 $E^{\rm miss}_T/\textbf{p}^{\ell}_{\rm T}>$ & 1.9 (2.8)  &  -  & -  & 1.0  &  -  & -   \\
$\Delta\phi_{\rm min}(\textbf{p}_{\rm T}(j_i),\textbf{p}^{\rm miss}_{\rm T})~\text{[rad]} >$ & -  & 0.85 (1.5)  &  -  & -  & 0.85 (1.5)  &  - \\
$p_T(\ell_2)~\text{[GeV]} >$ & -  &  -  & 190 (370) & -  &  -  & 190 (300)   \\
$M_{T2}~\text{[GeV]} >$ & - (140)  &  - (120)  & 200 (300) & 130 (230)   & 100  & 200 (300)  \\
DNN response $>$   & 0.95  & 0.95 & 0.95 & 0.95  & 0.95 & 0.95 \\
\hline\hline
\end{tabular}}
\label{tab5}
\end{table}

We present in Table~\ref{tab5} the set of preselection criteria as well as the analysis cuts used on the signal and background events.  The cuts are optimized depending on the category the benchmarks belong to and the center-of-mass energy. 

\begin{figure}[H]
\centering
\includegraphics[width=0.49\textwidth]{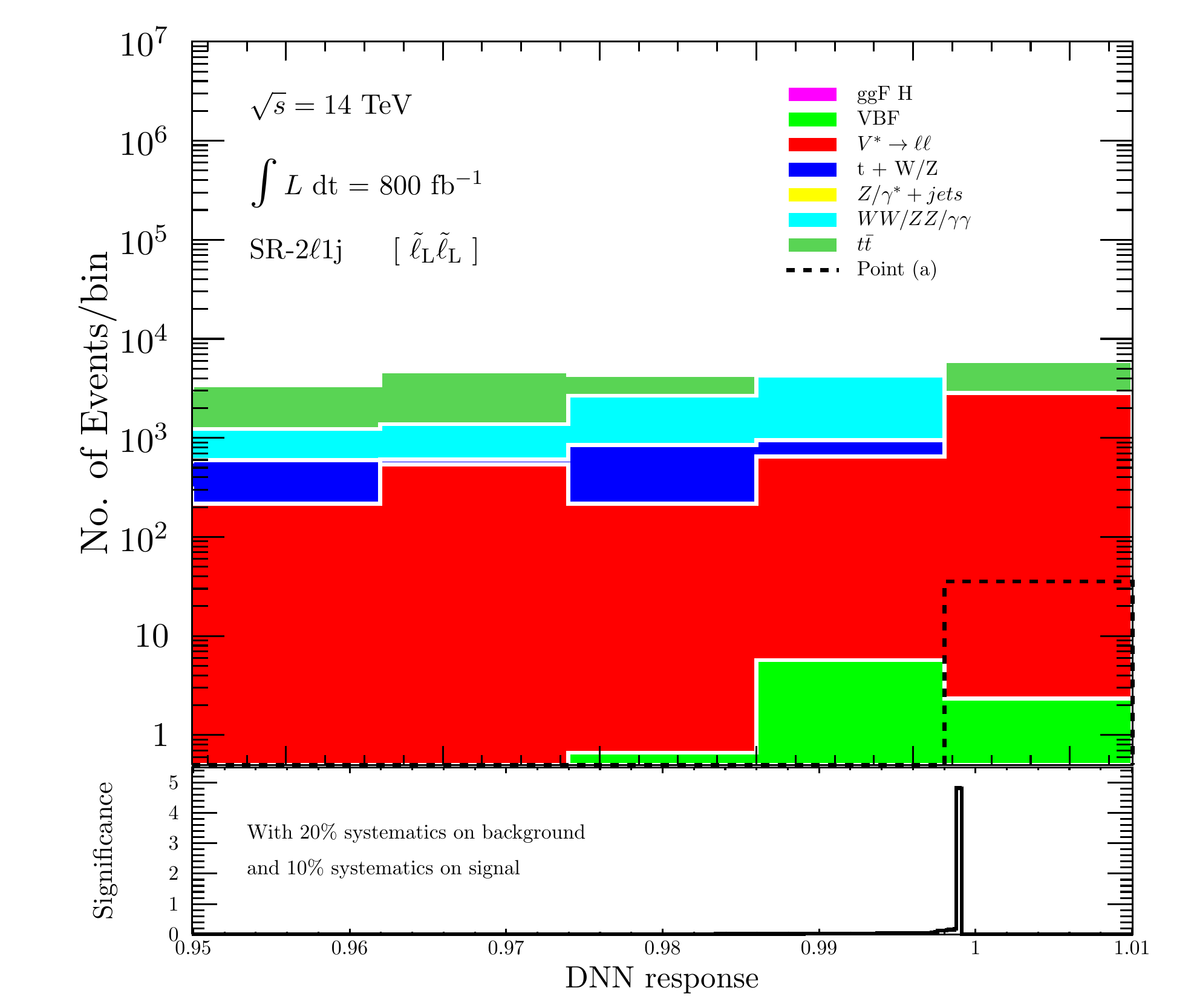}
\includegraphics[width=0.49\textwidth]{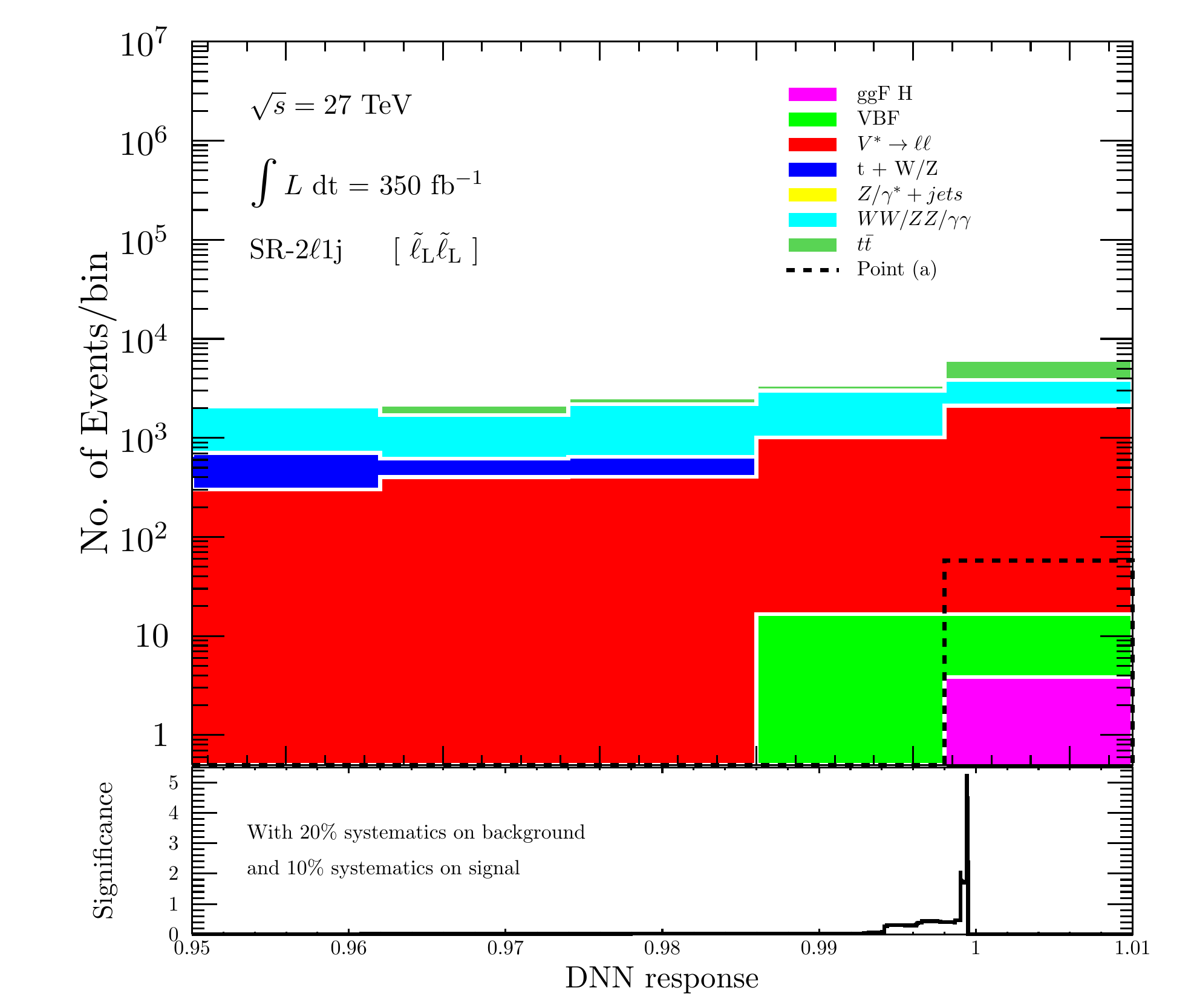} \\
\includegraphics[width=0.49\textwidth]{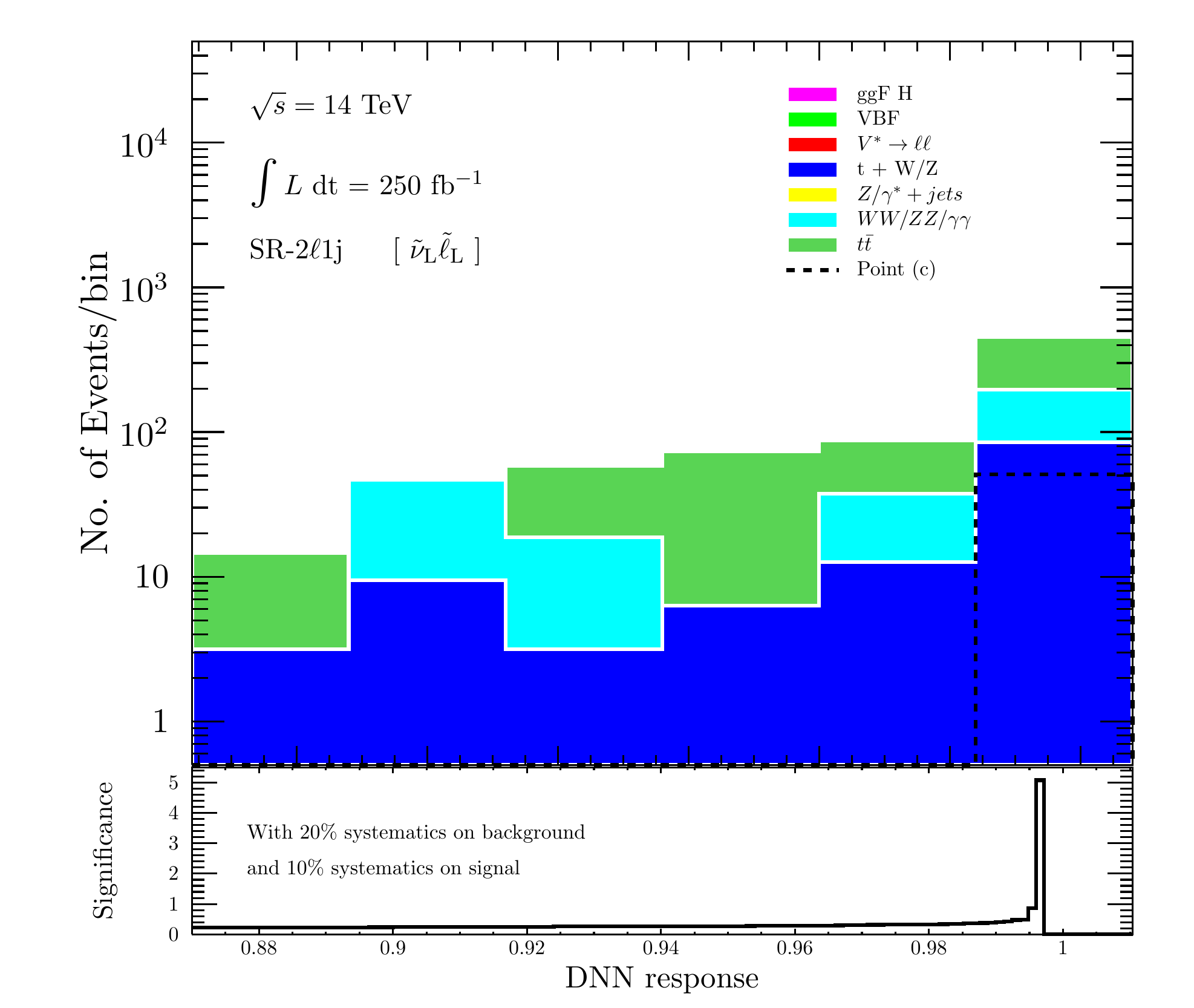}
\includegraphics[width=0.49\textwidth]{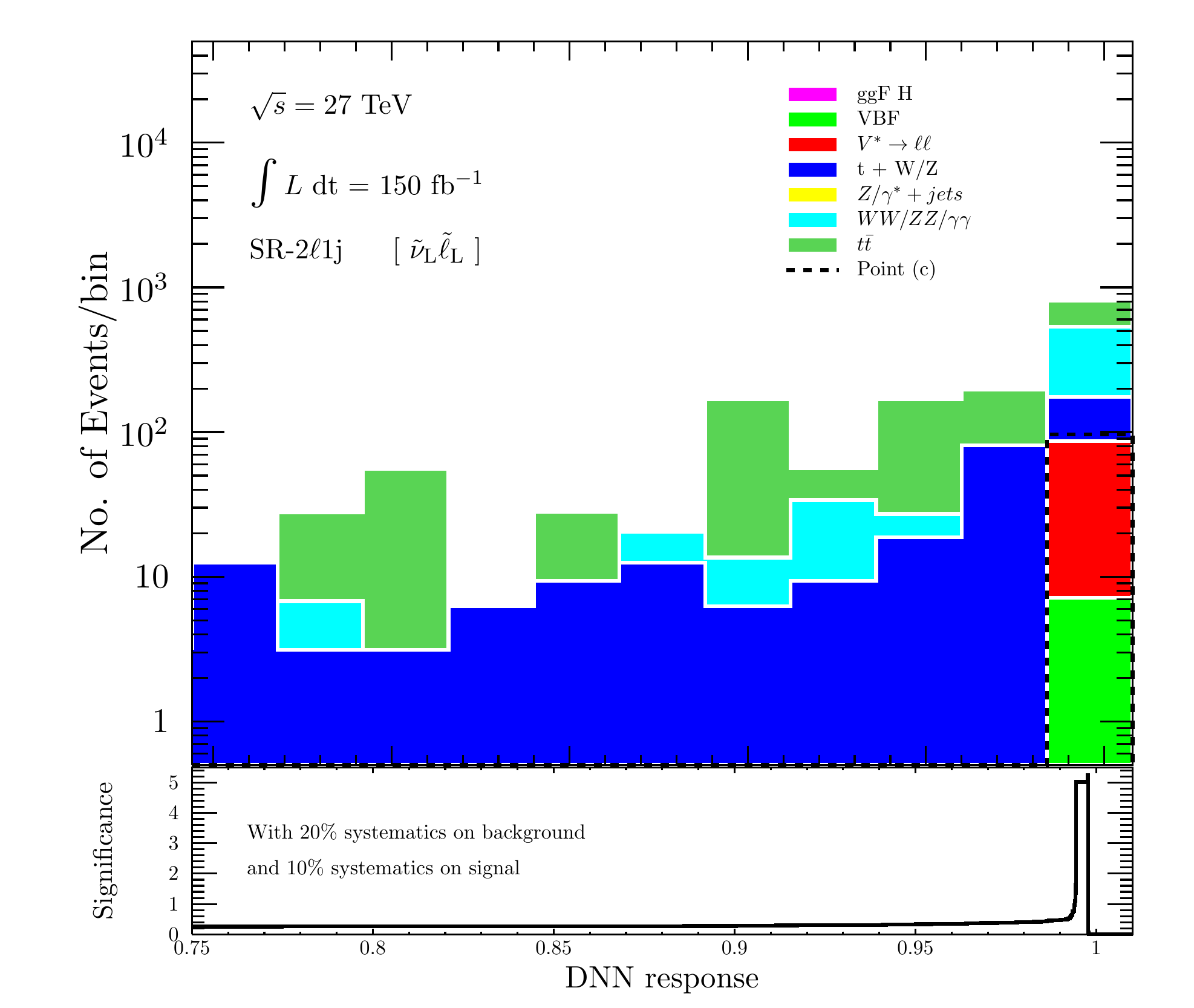}
\caption{Distributions in the `DNN response' variable for benchmark (a) (top panels) and benchmark (c) (bottom panels) at 14 TeV (left) and 27 TeV (right). A distribution in the signal significance as a function of the cut on the `DNN response' is shown in the bottom pads of each panel. } 
\label{dnn_dist}
\end{figure}

After training the DNN on the signal and background events and making sure no overtraining has occurred, the testing phase is done on a separate set of signal and background events. The end result is the construction of a new discriminating variable called the `DNN response' which tends to take values closer to one for the signal. Signal (S) and background (B) distributions in the `DNN response' variable are shown in Fig.~\ref{dnn_dist} for benchmarks (a) (top panels) and (c) (bottom panels) at 14 TeV and 27 TeV. In the bottom pad of each panel, we show the distribution in the signal significance $Z$ as a function of the cut on `DNN response' and taking into consideration the systematics, i.e.,
\begin{align}
Z=\frac{S}{\sqrt{S+B+(\delta_S S)^2+(\delta_B B)^2}},
\label{significance}
\end{align}
where $\delta_S$ and $\delta_B$ are the systematic uncertainties in the signal and background estimates. The recommendations on systematic uncertainties (known as `YR18' uncertainties) published in the CERN's yellow reports~\cite{CidVidal:2018eel,Cepeda:2019klc} suggest an overall 20\% uncertainty in the background and 10\% in the SUSY signal. Notice that a $5\sigma$ value is reached at the integrated luminosities shown which can be attained at HL-LHC and HE-LHC.

Of the two signal regions considered, we find that the most optimal one is the single jet signal region, which has also been shown to be true in previous LHC searches~\cite{Aad:2019vnb}. This is evident from Table~\ref{tab6} where the integrated luminosity for discovery of the benchmarks in SR-$2\ell2$j is much larger than in SR-$2\ell1$j after combining all production channels. The presented integrated luminosities include the effect of systematics.

\begin{table}[H]
\centering
\caption{\label{tab6} The estimated integrated luminosities, in fb$^{-1}$, for discovery of benchmarks of Table~\ref{tab1} at 14 TeV and 27 TeV after combining all production channels and including systematics in the signal and background. } 
\resizebox{0.75\textwidth}{!}{\begin{tabular}{|ccc|cc|}
\hline\hline
Model & \multicolumn{2}{c|}{SR-$2\ell1$j}& \multicolumn{2}{c|}{SR-$2\ell2$j}  \\
\hline \rule{0pt}{3ex}
& $\mathcal{L}$ at 14 TeV & $\mathcal{L}$ at 27 TeV & $\mathcal{L}$ at 14 TeV & $\mathcal{L}$ at 27 TeV\\
\hline \rule{0pt}{3ex}
\!\!(a) & 880 & 310 & 1262 & 694  \\
(b) & 200 & 50 & 1860 & 715  \\
(c) & 148 & 75 & 1887 & 1320  \\
(d) & 425 & 252 & $\cdots$ &  2804  \\
(e) & 1040 & 232 & 1738 & 1194  \\
(f) & 730 & 152 & 2074 & 689  \\
(g) & 970 & 202 & $\cdots$ & 1031  \\
 \hline\hline
\end{tabular}}
\end{table}

\begin{figure}[H]
\centering
\includegraphics[width=0.75\textwidth]{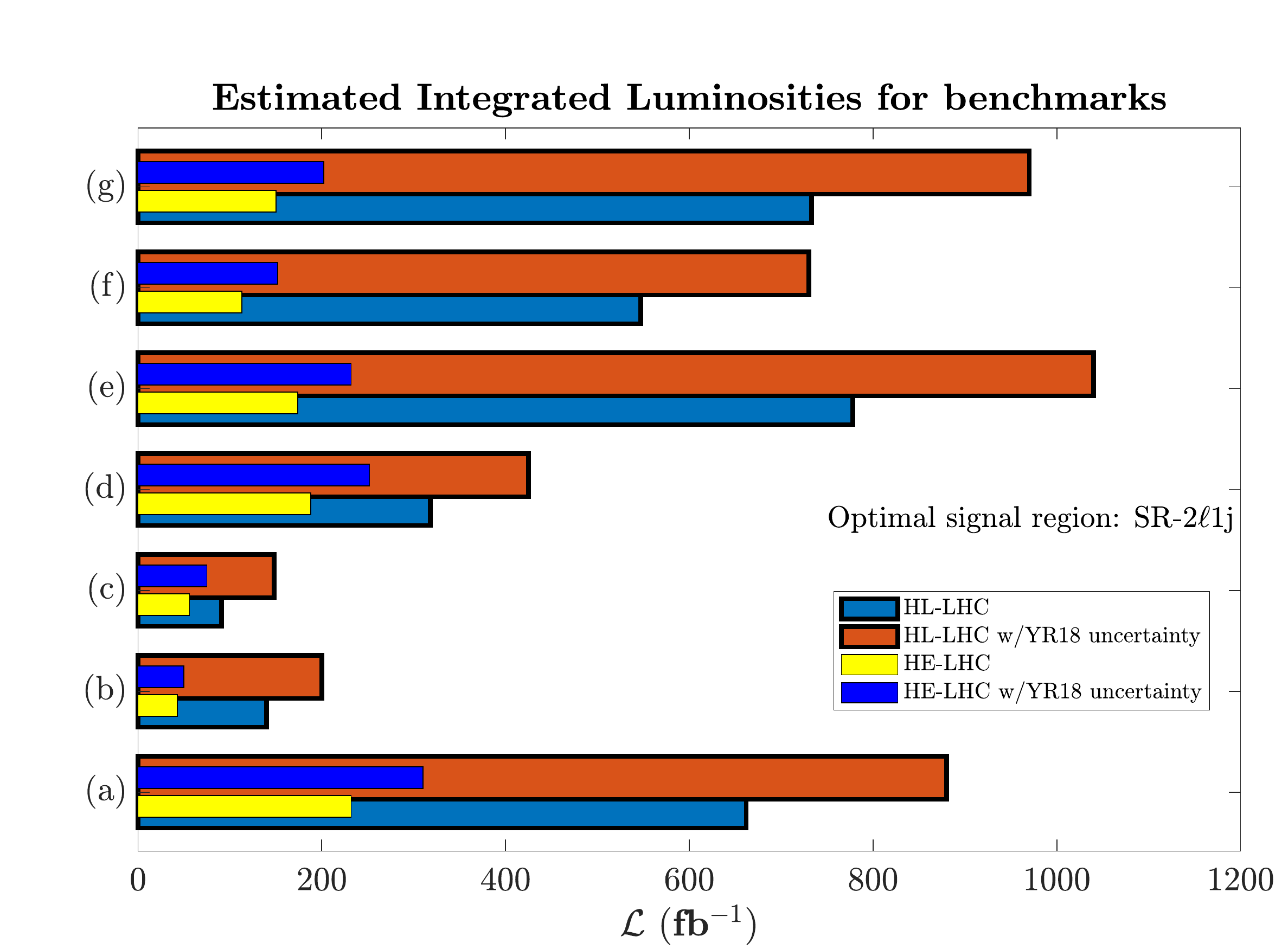}
\caption{The integrated luminosities, $\mathcal{L}$, needed for discovery of  SUSY at HL-LHC and HE-LHC assuming that \amu~arises from SUSY loops. Values of $\mathcal{L}$ are shown before and after including the `YR18' uncertainties on the signal and background. } 
\label{figlumi}
\end{figure}

As a comparison between HL-LHC and HE-LHC and to see the effect of systematic uncertainties, we plot the estimated integrated luminosities for discovery of benchmarks (a)$-$(g) at 14 TeV and 27 TeV in Fig.~\ref{figlumi}. Benchmarks (b) and (c) require 200 fb$^{-1}$ and 148 fb$^{-1}$ at 14 TeV which should be attained in the coming run of LHC. The rest require more than  400 fb$^{-1}$ but are all within the reach of HL-LHC. The same benchmarks require much smaller integrated luminosities for discovery at HE-LHC. 

We note that recently several works, including those within SUSY,  have come out regarding an explanation of the Fermilab muon
anomaly~\cite{Iwamoto:2021aaf,Gu:2021mjd,VanBeekveld:2021tgn,Yin:2021mls,Wang:2021bcx,Cao:2021tuh,Chakraborti:2021dli,Cox:2021gqq,Han:2021ify,Baum:2021qzx,Ahmed:2021htr,Baer:2021aax,Endo:2021zal,Ibe:2021cvf,Chakraborti:2021bmv,Anchordoqui:2021llp,Athron:2021iuf,Lu:2021vcp,Li:2021lnz,Cen:2021iwv,Borah:2021jzu,Zhou:2021vnf,Altmannshofer:2021hfu,Dasgupta:2021dnl,Jueid:2021avn,Carpio:2021jhu,Babu:2021jnu,Zheng:2021wnu,Zhang:2021dgl,De:2021crr,Jeong:2021qey,Yu:2021suw,Ellis:2021zmg,Frank:2021nkq,Abdughani:2021pdc,Abdughani:2019wai}.

\section{Conclusion \label{6}}

       The recent results from the Fermilab Collaboration confirm the earlier result from the Brookhaven 
       experiment regarding a deviation from the Standard Model prediction of the muon $g-2$ anomaly.
       Specifically, the combined Fermilab and Brookhaven results give a $4.1\sigma$ deviation from the
       Standard Model prediction compared to $3.7\sigma$ prediction for Brookhaven which both supports
       the observation  and further strengthens it. 
       
 In this paper we have pointed to the implications of the Fermilab result for SUSY discovery at the LHC.
       To this end we have analyzed  the effect of the Fermilab constraint on the SUGRA parameter space. This is done in a
       comprehensive way with the help of neural networks where we have used the constraints of the Higgs boson mass
       and the relic density along with the muon $g-2$ to delineate the region of the parameter space
       consistent with those constraints. The neural network analysis identifies the allowed region as 
       the one that is generated by gluino-driven radiative breaking of the electroweak symmetry.  This region of the 
       parameter space naturally leads to a split sparticle spectrum constituted of light sleptons and light weakinos but
       a heavy gluino and heavy squarks which allows one to satisfy muon $g-2$ constraint and the Higgs boson constraint
       within a high scale model.  We emphasize the existence of a light spectrum and their possible early discovery 
       at  the LHC. We propose a set of benchmarks relevant for Snowmass study related to the energy frontier. 
       Further, we demonstrate that 
       a deep neural network with  the `DNN response' variable constitutes a powerful discriminant to reject the Standard Model background which aids in the discovery of new physics signals. We exhibit the importance of this technique 
       by analyzing the integrated luminosities needed for the discovery of the 
        benchmarks, and show that some of the benchmarks can be discovered 
          at  HL-LHC with integrated luminosities as low as 150$-$300 fb$^{-1}$ while much smaller integrated luminosities are required at HE-LHC.

 We propose that it would be fruitful if the future SUSY analyses at the LHC are carried out with the inclusion of the Fermilab $g-2$ constraint  and further such analyses need to
           go beyond generic simplified models to well motivated high scale models. The basic reason for such analyses 
           is that in well motivated models the branching ratio to any production channel is almost never unity as generally 
           assumed in simplified models.          
               Thus dedicated studies for the production and signal analysis of sleptons and weakinos at the LHC
          within Snowmass study are timely.   \\~\\

 The research of AA and MK was supported by the BMBF under contract 05H18PMCC1. The research of PN  was supported in part by the NSF Grant PHY-1913328.


\newpage

\begin{thebibliography}{999}

\bibitem{Aoyama:2020ynm}
T.~Aoyama, N.~Asmussen, M.~Benayoun, J.~Bijnens, T.~Blum, M.~Bruno, I.~Caprini, C.~M.~Carloni Calame, M.~C\`e and G.~Colangelo, \textit{et al.}
Phys. Rept. \textbf{887}, 1-166 (2020)
doi:10.1016/j.physrep.2020.07.006
[arXiv:2006.04822 [hep-ph]].

\bibitem{Davier:2019can}
M.~Davier, A.~Hoecker, B.~Malaescu and Z.~Zhang,
Eur. Phys. J. C \textbf{80}, no.3, 241 (2020)
[erratum: Eur. Phys. J. C \textbf{80}, no.5, 410 (2020)]
doi:10.1140/epjc/s10052-020-7792-2
[arXiv:1908.00921 [hep-ph]].

\bibitem{Davier:2017zfy}
M.~Davier, A.~Hoecker, B.~Malaescu and Z.~Zhang,
Eur. Phys. J. C \textbf{77}, no.12, 827 (2017)
doi:10.1140/epjc/s10052-017-5161-6
[arXiv:1706.09436 [hep-ph]].

\bibitem{Davier:2010nc}
M.~Davier, A.~Hoecker, B.~Malaescu and Z.~Zhang,
Eur. Phys. J. C \textbf{71}, 1515 (2011)
[erratum: Eur. Phys. J. C \textbf{72}, 1874 (2012)]
doi:10.1140/epjc/s10052-012-1874-8
[arXiv:1010.4180 [hep-ph]].

\bibitem{Crivellin:2020zul}
A.~Crivellin, M.~Hoferichter, C.~A.~Manzari and M.~Montull,
Phys. Rev. Lett. \textbf{125}, no.9, 091801 (2020)
doi:10.1103/PhysRevLett.125.091801
[arXiv:2003.04886 [hep-ph]].

\bibitem{Keshavarzi:2020bfy}
A.~Keshavarzi, W.~J.~Marciano, M.~Passera and A.~Sirlin,
Phys. Rev. D \textbf{102}, no.3, 033002 (2020)
doi:10.1103/PhysRevD.102.033002
[arXiv:2006.12666 [hep-ph]].

\bibitem{Colangelo:2020lcg}
G.~Colangelo, M.~Hoferichter and P.~Stoffer,
Phys. Lett. B \textbf{814}, 136073 (2021)
doi:10.1016/j.physletb.2021.136073
[arXiv:2010.07943 [hep-ph]].

\bibitem{Abi:2021gix}
B.~Abi \textit{et al.} [Muon g-2],
Phys. Rev. Lett. \textbf{126}, no.14, 141801 (2021)
doi:10.1103/PhysRevLett.126.141801
[arXiv:2104.03281 [hep-ex]].

\bibitem{Bennett:2006fi}
G.~W.~Bennett \textit{et al.} [Muon g-2],
Phys. Rev. D \textbf{73}, 072003 (2006)
doi:10.1103/PhysRevD.73.072003
[arXiv:hep-ex/0602035 [hep-ex]].

\bibitem{Tanabishi}
M. Tanabashi, et al., Review of Particle Physics, Phys. Rev. D98 (3) (2018) 030001. 
doi:10.1103/PhysRevD.98. 030001

\bibitem{Borsanyi:2020mff}
S.~Borsanyi, Z.~Fodor, J.~N.~Guenther, C.~Hoelbling, S.~D.~Katz, L.~Lellouch, T.~Lippert, K.~Miura, L.~Parato and K.~K.~Szabo, \textit{et al.}
Nature \textbf{593}, no.7857, 51-55 (2021)
doi:10.1038/s41586-021-03418-1
[arXiv:2002.12347 [hep-lat]].

\bibitem{Aboubrahim:2021rwz}
A.~Aboubrahim, M.~Klasen and P.~Nath,
[arXiv:2104.03839 [hep-ph]]. (To appear in PRD).

\bibitem{Kosower:1983yw}
D.~A.~Kosower, L.~M.~Krauss and N.~Sakai,
Phys. Lett. B \textbf{133}, 305-310 (1983)
doi:10.1016/0370-2693(83)90152-1

\bibitem{Yuan:1984ww}
T.~C.~Yuan, R.~L.~Arnowitt, A.~H.~Chamseddine and P.~Nath,
Z. Phys. C \textbf{26}, 407 (1984)
doi:10.1007/BF01452567

\bibitem{Lopez:1993vi}
J.~L.~Lopez, D.~V.~Nanopoulos and X.~Wang,
Phys. Rev. D \textbf{49}, 366-372 (1994)
doi:10.1103/PhysRevD.49.366
[arXiv:hep-ph/9308336 [hep-ph]].

\bibitem{Chattopadhyay:1995ae}
U.~Chattopadhyay and P.~Nath,
Phys. Rev. D \textbf{53}, 1648-1657 (1996)
doi:10.1103/PhysRevD.53.1648
[arXiv:hep-ph/9507386 [hep-ph]].

\bibitem{Moroi:1995yh}
T.~Moroi,
Phys. Rev. D \textbf{53}, 6565-6575 (1996)
[erratum: Phys. Rev. D \textbf{56}, 4424 (1997)]
doi:10.1103/PhysRevD.53.6565
[arXiv:hep-ph/9512396 [hep-ph]].

\bibitem{Aad:2012tfa}
G.~Aad \textit{et al.} [ATLAS],
Phys. Lett. B \textbf{716}, 1-29 (2012)
doi:10.1016/j.physletb.2012.08.020
[arXiv:1207.7214 [hep-ex]].

\bibitem{Chatrchyan:2012ufa}
S.~Chatrchyan \textit{et al.} [CMS],
Phys. Lett. B \textbf{716}, 30-61 (2012)
doi:10.1016/j.physletb.2012.08.021
[arXiv:1207.7235 [hep-ex]].

\bibitem{Akula:2011aa}
S.~Akula, B.~Altunkaynak, D.~Feldman, P.~Nath and G.~Peim,
Phys. Rev. D \textbf{85}, 075001 (2012)
doi:10.1103/PhysRevD.85.075001
[arXiv:1112.3645 [hep-ph]].

\bibitem{higgs7tev1}
    A.~Arbey, M.~Battaglia, A.~Djouadi, F.~Mahmoudi and J.~Quevillon,
  Phys.\ Lett.\ B {\bf 708}, 162 (2012);
  H.~Baer, V.~Barger and A.~Mustafayev,
  Phys.\ Rev.\ D {\bf 85}, 075010 (2012);
      J.~Ellis   and K.~A.~Olive,
  Eur.\ Phys.\ J.\ C {\bf 72}, 2005 (2012); 
    S.~Heinemeyer, O.~Stal and G.~Weiglein,
  Phys.\ Lett.\ B {\bf 710}, 201 (2012);
  
\bibitem{Akula:2013ioa}
S.~Akula and P.~Nath,
Phys. Rev. D \textbf{87}, no.11, 115022 (2013)
doi:10.1103/PhysRevD.87.115022
[arXiv:1304.5526 [hep-ph]].

\bibitem{Aboubrahim:2019vjl}
A.~Aboubrahim and P.~Nath,
Phys. Rev. D \textbf{100}, no.1, 015042 (2019)
doi:10.1103/PhysRevD.100.015042
[arXiv:1905.04601 [hep-ph]].

\bibitem{Aboubrahim:2020dqw}
A.~Aboubrahim, P.~Nath and R.~M.~Syed,
JHEP \textbf{01}, 047 (2021)
doi:10.1007/JHEP01(2021)047
[arXiv:2005.00867 [hep-ph]].

\bibitem{Aboubrahim:2021phn}
A.~Aboubrahim, P.~Nath and R.~M.~Syed,
JHEP \textbf{06}, 002 (2021)
doi:10.1007/JHEP06(2021)002
[arXiv:2104.10114 [hep-ph]].

\bibitem{Hollingsworth:2021sii}
J.~Hollingsworth, M.~Ratz, P.~Tanedo and D.~Whiteson,
[arXiv:2103.06957 [hep-th]].

\bibitem{Balazs:2021uhg}
C.~Bal\'azs \textit{et al.} [DarkMachines High Dimensional Sampling Group],
JHEP \textbf{05}, 108 (2021)
doi:10.1007/JHEP05(2021)108
[arXiv:2101.04525 [hep-ph]].

\bibitem{sugrauni}
A.~H.~Chamseddine, R.~Arnowitt and P.~Nath,
  Phys.\ Rev.\ Lett.\  {\bf 49} (1982) 970;
  P.~Nath, R.~L.~Arnowitt and A.~H.~Chamseddine,
  Nucl.\ Phys.\  B {\bf 227}, 121 (1983);
 L.~J.~Hall, J.~D.~Lykken and S.~Weinberg,
  Phys.\ Rev.\ D {\bf 27}, 2359 (1983).
  doi:10.1103/PhysRevD.27.2359
  
\bibitem{Ellis:1985jn}
J.~R.~Ellis, K.~Enqvist, D.~V.~Nanopoulos and K.~Tamvakis,
Phys. Lett. B \textbf{155}, 381-386 (1985)
doi:10.1016/0370-2693(85)91591-6

\bibitem{Feldman:2009zc}
D.~Feldman, Z.~Liu and P.~Nath,
Phys. Rev. D \textbf{80}, 015007 (2009)
doi:10.1103/PhysRevD.80.015007
[arXiv:0905.1148 [hep-ph]].

\bibitem{Belyaev:2018vkl}
A.~S.~Belyaev, S.~F.~King and P.~B.~Schaefers,
Phys. Rev. D \textbf{97}, no.11, 115002 (2018)
doi:10.1103/PhysRevD.97.115002
[arXiv:1801.00514 [hep-ph]].

\bibitem{nonuni2}
 A.~Corsetti and P.~Nath,
  Phys.\ Rev.\  D {\bf 64}, 125010 (2001);
  A.~Birkedal-Hansen and B.~D.~Nelson,
  Phys.\ Rev.\  D {\bf 67}, 095006 (2003);
  G.~Belanger, F.~Boudjema, A.~Cottrant, A.~Pukhov and A.~Semenov,
  Nucl.\ Phys.\  B {\bf 706}, 411 (2005);
  H.~Baer, A.~Mustafayev, E.~K.~Park, S.~Profumo and X.~Tata,
JHEP \textbf{04} (2006), 041
doi:10.1088/1126-6708/2006/04/041
[arXiv:hep-ph/0603197 [hep-ph]];
  I.~Gogoladze, F.~Nasir, Q.~Shafi and C.~S.~Un,
  Phys.\ Rev.\ D {\bf 90}, no. 3, 035008 (2014)
  doi:10.1103/PhysRevD.90.035008;
  S.~P.~Martin,
  Phys.\ Rev.\ D {\bf 79}, 095019 (2009)
  doi:10.1103/PhysRevD.79.095019
  
\bibitem{Masiero:1982fe}
A.~Masiero, D.~V.~Nanopoulos, K.~Tamvakis and T.~Yanagida,
Phys. Lett. B \textbf{115}, 380-384 (1982)
doi:10.1016/0370-2693(82)90522-6

\bibitem{Grinstein:1982um}
B.~Grinstein,
Nucl. Phys. B \textbf{206}, 387 (1982)
doi:10.1016/0550-3213(82)90275-9

\bibitem{Babu:2006nf}
K.~S.~Babu, I.~Gogoladze and Z.~Tavartkiladze,
Phys. Lett. B \textbf{650}, 49-56 (2007)
doi:10.1016/j.physletb.2007.02.050
[arXiv:hep-ph/0612315 [hep-ph]].

\bibitem{Babu:2011tw}
K.~S.~Babu, I.~Gogoladze, P.~Nath and R.~M.~Syed,
Phys. Rev. D \textbf{85}, 075002 (2012)
doi:10.1103/PhysRevD.85.075002
[arXiv:1112.5387 [hep-ph]].

\bibitem{Nath:2006ut}
P.~Nath and P.~Fileviez Perez,
Phys. Rept. \textbf{441}, 191-317 (2007)
doi:10.1016/j.physrep.2007.02.010
[arXiv:hep-ph/0601023 [hep-ph]].

\bibitem{Ananthanarayan:1992cd}
B.~Ananthanarayan, G.~Lazarides and Q.~Shafi,
Phys. Lett. B \textbf{300}, 245-250 (1993)
doi:10.1016/0370-2693(93)90361-K

\bibitem{Chattopadhyay:2001va}
U.~Chattopadhyay, A.~Corsetti and P.~Nath,
Phys. Rev. D \textbf{66}, 035003 (2002)
doi:10.1103/PhysRevD.66.035003
[arXiv:hep-ph/0201001 [hep-ph]].

\bibitem{Ibrahim:1999aj}
T.~Ibrahim and P.~Nath,
Phys. Rev. D \textbf{62}, 015004 (2000)
doi:10.1103/PhysRevD.62.015004
[arXiv:hep-ph/9908443 [hep-ph]].

\bibitem{Ibrahim:2001ym}
T.~Ibrahim, U.~Chattopadhyay and P.~Nath,
Phys. Rev. D \textbf{64}, 016010 (2001)
doi:10.1103/PhysRevD.64.016010
[arXiv:hep-ph/0102324 [hep-ph]].

\bibitem{Aboubrahim:2016xuz}
A.~Aboubrahim, T.~Ibrahim and P.~Nath,
Phys. Rev. D \textbf{94}, no.1, 015032 (2016)
doi:10.1103/PhysRevD.94.015032
[arXiv:1606.08336 [hep-ph]].

\bibitem{Ibrahim:1998je}
T.~Ibrahim and P.~Nath,
Phys. Rev. D \textbf{58}, 111301 (1998)
[erratum: Phys. Rev. D \textbf{60}, 099902 (1999)]
doi:10.1103/PhysRevD.58.111301
[arXiv:hep-ph/9807501 [hep-ph]].

\bibitem{Chattopadhyay:1998wb}
U.~Chattopadhyay, T.~Ibrahim and P.~Nath,
Phys. Rev. D \textbf{60}, 063505 (1999)
doi:10.1103/PhysRevD.60.063505
[arXiv:hep-ph/9811362 [hep-ph]].

\bibitem{Ibrahim:2000tx}
T.~Ibrahim and P.~Nath,
Phys. Rev. D \textbf{62}, 095001 (2000)
doi:10.1103/PhysRevD.62.095001
[arXiv:hep-ph/0004098 [hep-ph]].

\bibitem{Debove:2011xj}
J.~Debove, B.~Fuks and M.~Klasen,
Nucl. Phys. B \textbf{849}, 64-79 (2011)
doi:10.1016/j.nuclphysb.2011.03.015
[arXiv:1102.4422 [hep-ph]].

\bibitem{Fuks:2013vua}
B.~Fuks, M.~Klasen, D.~R.~Lamprea and M.~Rothering,
Eur. Phys. J. C \textbf{73}, 2480 (2013)
doi:10.1140/epjc/s10052-013-2480-0
[arXiv:1304.0790 [hep-ph]].

\bibitem{Alwall:2014hca}
J.~Alwall, R.~Frederix, S.~Frixione, V.~Hirschi, F.~Maltoni, O.~Mattelaer, H.~S.~Shao, T.~Stelzer, P.~Torrielli and M.~Zaro,
JHEP \textbf{07}, 079 (2014)
doi:10.1007/JHEP07(2014)079
[arXiv:1405.0301 [hep-ph]].

\bibitem{Sjostrand:2014zea}
T.~Sj\"ostrand, S.~Ask, J.~R.~Christiansen, R.~Corke, N.~Desai, P.~Ilten, S.~Mrenna, S.~Prestel, C.~O.~Rasmussen and P.~Z.~Skands,
Comput. Phys. Commun. \textbf{191}, 159-177 (2015)
doi:10.1016/j.cpc.2015.01.024
[arXiv:1410.3012 [hep-ph]].

\bibitem{deFavereau:2013fsa}
J.~de Favereau \textit{et al.} [DELPHES 3],
JHEP \textbf{02}, 057 (2014)
doi:10.1007/JHEP02(2014)057
[arXiv:1307.6346 [hep-ex]].

\bibitem{Speckmayer:2010zz}
P.~Speckmayer, A.~Hocker, J.~Stelzer and H.~Voss,
J. Phys. Conf. Ser. \textbf{219}, 032057 (2010)
doi:10.1088/1742-6596/219/3/032057

\bibitem{Antcheva:2011zz}
I.~Antcheva, M.~Ballintijn, B.~Bellenot, M.~Biskup, R.~Brun, N.~Buncic, P.~Canal, D.~Casadei, O.~Couet and V.~Fine, \textit{et al.}
Comput. Phys. Commun. \textbf{182}, 1384-1385 (2011)
doi:10.1016/j.cpc.2011.02.008

\bibitem{CidVidal:2018eel}
X.~Cid Vidal, M.~D'Onofrio, P.~J.~Fox, R.~Torre, K.~A.~Ulmer, A.~Aboubrahim, A.~Albert, J.~Alimena, B.~C.~Allanach and C.~Alpigiani, \textit{et al.}
CERN Yellow Rep. Monogr. \textbf{7}, 585-865 (2019)
doi:10.23731/CYRM-2019-007.585
[arXiv:1812.07831 [hep-ph]].

\bibitem{Cepeda:2019klc}
M.~Cepeda, S.~Gori, P.~Ilten, M.~Kado, F.~Riva, R.~Abdul Khalek, A.~Aboubrahim, J.~Alimena, S.~Alioli and A.~Alves, \textit{et al.}
CERN Yellow Rep. Monogr. \textbf{7}, 221-584 (2019)
doi:10.23731/CYRM-2019-007.221
[arXiv:1902.00134 [hep-ph]].

\bibitem{Aad:2019vnb}
G.~Aad \textit{et al.} [ATLAS],
Eur. Phys. J. C \textbf{80}, no.2, 123 (2020)
doi:10.1140/epjc/s10052-019-7594-6
[arXiv:1908.08215 [hep-ex]].

\bibitem{Iwamoto:2021aaf}
S.~Iwamoto, T.~T.~Yanagida and N.~Yokozaki,
[arXiv:2104.03223 [hep-ph]].

\bibitem{Gu:2021mjd}
Y.~Gu, N.~Liu, L.~Su and D.~Wang,
[arXiv:2104.03239 [hep-ph]].

\bibitem{VanBeekveld:2021tgn}
M.~Van Beekveld, W.~Beenakker, M.~Schutten and J.~De Wit,
[arXiv:2104.03245 [hep-ph]].

\bibitem{Yin:2021mls}
W.~Yin,
JHEP \textbf{06}, 029 (2021)
doi:10.1007/JHEP06(2021)029
[arXiv:2104.03259 [hep-ph]].

\bibitem{Wang:2021bcx}
F.~Wang, L.~Wu, Y.~Xiao, J.~M.~Yang and Y.~Zhang,
[arXiv:2104.03262 [hep-ph]].

\bibitem{Cao:2021tuh}
J.~Cao, J.~Lian, Y.~Pan, D.~Zhang and P.~Zhu,
[arXiv:2104.03284 [hep-ph]].

\bibitem{Chakraborti:2021dli}
M.~Chakraborti, S.~Heinemeyer and I.~Saha,
[arXiv:2104.03287 [hep-ph]].

\bibitem{Cox:2021gqq}
P.~Cox, C.~Han and T.~T.~Yanagida,
[arXiv:2104.03290 [hep-ph]].

\bibitem{Han:2021ify}
C.~Han,
[arXiv:2104.03292 [hep-ph]].

\bibitem{Baum:2021qzx}
S.~Baum, M.~Carena, N.~R.~Shah and C.~E.~M.~Wagner,
[arXiv:2104.03302 [hep-ph]].

\bibitem{Ahmed:2021htr}
W.~Ahmed, I.~Khan, J.~Li, T.~Li, S.~Raza and W.~Zhang,
[arXiv:2104.03491 [hep-ph]].

\bibitem{Baer:2021aax}
H.~Baer, V.~Barger and H.~Serce,
Phys. Lett. B \textbf{820}, 136480 (2021)
doi:10.1016/j.physletb.2021.136480
[arXiv:2104.07597 [hep-ph]].

\bibitem{Endo:2021zal}
M.~Endo, K.~Hamaguchi, S.~Iwamoto and T.~Kitahara,
[arXiv:2104.03217 [hep-ph]].

\bibitem{Ibe:2021cvf}
M.~Ibe, S.~Kobayashi, Y.~Nakayama and S.~Shirai,
[arXiv:2104.03289 [hep-ph]].

\bibitem{Chakraborti:2021bmv}
M.~Chakraborti, L.~Roszkowski and S.~Trojanowski,
JHEP \textbf{05}, 252 (2021)
doi:10.1007/JHEP05(2021)252
[arXiv:2104.04458 [hep-ph]].

\bibitem{Anchordoqui:2021llp}
L.~A.~Anchordoqui, I.~Antoniadis, X.~Huang, D.~Lust and T.~R.~Taylor,
doi:10.1002/prop.202100084
[arXiv:2104.06854 [hep-ph]].

\bibitem{Athron:2021iuf}
P.~Athron, C.~Bal\'azs, D.~H.~Jacob, W.~Kotlarski, D.~St\"ockinger and H.~St\"ockinger-Kim,
[arXiv:2104.03691 [hep-ph]].

\bibitem{Lu:2021vcp}
C.~T.~Lu, R.~Ramos and Y.~L.~Sming Tsai,
[arXiv:2104.04503 [hep-ph]].

\bibitem{Li:2021lnz}
T.~Li, M.~A.~Schmidt, C.~Y.~Yao and M.~Yuan,
[arXiv:2104.04494 [hep-ph]].

\bibitem{Cen:2021iwv}
J.~Y.~Cen, Y.~Cheng, X.~G.~He and J.~Sun,
[arXiv:2104.05006 [hep-ph]].

\bibitem{Borah:2021jzu}
D.~Borah, M.~Dutta, S.~Mahapatra and N.~Sahu,
[arXiv:2104.05656 [hep-ph]].

\bibitem{Zhou:2021vnf}
S.~Zhou,
[arXiv:2104.06858 [hep-ph]].

\bibitem{Altmannshofer:2021hfu}
W.~Altmannshofer, S.~A.~Gadam, S.~Gori and N.~Hamer,
[arXiv:2104.08293 [hep-ph]].

\bibitem{Dasgupta:2021dnl}
A.~Dasgupta, S.~K.~Kang and M.~Park,
[arXiv:2104.09205 [hep-ph]].

\bibitem{Jueid:2021avn}
A.~Jueid, J.~Kim, S.~Lee and J.~Song,
[arXiv:2104.10175 [hep-ph]].

\bibitem{Carpio:2021jhu}
J.~A.~Carpio, K.~Murase, I.~M.~Shoemaker and Z.~Tabrizi,
[arXiv:2104.15136 [hep-ph]].

\bibitem{Babu:2021jnu}
K.~S.~Babu, S.~Jana, M.~Lindner and V.~P.~K,
[arXiv:2104.03291 [hep-ph]].

\bibitem{Frank:2021nkq}
M.~Frank, Y.~Hi\c{c}y\i{}lmaz, S.~Mondal, \"O.~\"Ozdal and C.~S.~\"Un,
[arXiv:2107.04116 [hep-ph]].

\bibitem{Ellis:2021zmg}
J.~Ellis, J.~L.~Evans, N.~Nagata, D.~V.~Nanopoulos and K.~A.~Olive,
[arXiv:2107.03025 [hep-ph]].

\bibitem{Yu:2021suw}
B.~Yu and S.~Zhou,
[arXiv:2106.11291 [hep-ph]].

\bibitem{Jeong:2021qey}
K.~S.~Jeong, J.~Kawamura and C.~B.~Park,
[arXiv:2106.04238 [hep-ph]].

\bibitem{De:2021crr}
B.~De, D.~Das, M.~Mitra and N.~Sahoo,
[arXiv:2106.00979 [hep-ph]].

\bibitem{Zhang:2021dgl}
D.~Zhang,
[arXiv:2105.08670 [hep-ph]].

\bibitem{Zheng:2021wnu}
M.~D.~Zheng and H.~H.~Zhang,
[arXiv:2105.06954 [hep-ph]].

\bibitem{Abdughani:2021pdc}
M.~Abdughani, Y.~Z.~Fan, L.~Feng, Y.~L.~S.~Tsai, L.~Wu and Q.~Yuan,
Sci. Bull. \textbf{66}, 2170-2174 (2021)
doi:10.1016/j.scib.2021.07.029
[arXiv:2104.03274 [hep-ph]].

\bibitem{Abdughani:2019wai}
M.~Abdughani, K.~I.~Hikasa, L.~Wu, J.~M.~Yang and J.~Zhao,
JHEP \textbf{11}, 095 (2019)
doi:10.1007/JHEP11(2019)095
[arXiv:1909.07792 [hep-ph]].



\end{thebibliography}
\end{document}